\documentclass[draft]{agujournal}

\draftfalse
\usepackage{url,hyperref}

\renewcommand{\textwidth}{15cm}
\usepackage{amsmath}

\journalname{JGR-Planets}

\newcommand{\mitbf}[1]{
  \hbox{\mathversion{bold}$#1$}}

\begin{document}

%
%


\title{Deviation of Mercury's spin axis from an exact Cassini state induced by dissipation}

%
%




\authors{Ian MacPherson \affil{1}, Mathieu Dumberry \affil{1}}
 
\affiliation{1}{Department of Physics, University of Alberta, Edmonton, Alberta, Canada.}






\correspondingauthor{Mathieu Dumberry}{dumberry@ualberta.ca}




\begin{keypoints}
\item Viscous and electromagnetic drag at the fluid core boundaries generate a deviation that does not exceed 0.1 arcsec
\item In units of arcsec, the phase lag from tidal dissipation follows the empirical relation (80/$Q$), where $Q$ is the quality factor
\item The maximum phase lag allowed by observations gives a lower limit on the bulk mantle viscosity of approximately $10^{17}$ Pa s
\end{keypoints}

%
%

\begin{abstract}
We compute predictions of the deviation of Mercury's spin axis from an exact Cassini state caused by tidal dissipation, and viscous and electromagnetic (EM) friction at the core-mantle boundary (CMB) and inner core boundary (ICB).  Viscous friction at the CMB generates a phase lead, viscous and EM friction at the ICB produce a phase lag; the magnitude of the deviation depends on the inner core size, kinematic viscosity and magnetic field strength, but cannot exceed an upper bound. For a small inner core, viscous friction at the CMB results in a maximum phase lead of 0.027 arcsec.  For a large inner core (radius $>1000$ km), EM friction at the ICB generates the largest phase lag, but it does not exceed 0.1 arcsec. Elastic deformations induced by the misaligned fluid and solid cores play a first order role in the phase lead/lag caused by viscous and EM coupling, and contribute to a perturbation in mantle obliquity on par with that caused by tidal deformations.  Tidal dissipation results in a phase lag and its magnitude (in units of arcsec) is given by the empirical relation (80/Q), where Q is the quality factor; Q=80 results in a phase lag of $\sim1$ arcsec.  A large inner core with a low viscosity of the order of $10^{17}$ Pa s or lower can significantly affect $Q$ and thus the resulting phase lag. The limited mantle phase lag suggested by observations ($<$10 arcsec) implies a lower limit on the bulk mantle viscosity of approximately $10^{17}$ Pa s.
\end{abstract}

\noindent{\bf Plain language summary:} As Mercury orbits the Sun, the plane of its orbit is slowly precessing about a fixed axis in space.  This locks the spin axis of Mercury into its own precession at the same rate.  This configuration is known as a Cassini state in which the spin axis is oriented in the same plane as that formed by the orbit normal and the fixed axis (the Cassini plane). Dissipation introduces a small deviation of Mercury's spin axis from the Cassini plane.  We compute predictions of this deviation.  We show that viscous and electromagnetic friction at the boundaries of the fluid core result in a limited deviation which does not exceed 0.1 arcsec.  Dissipation from tidal deformations produce a deviation that is inversely proportional to the mantle viscosity, a measure of how stiff the mantle is.  Measurements of the orientations of Mercury's spin axis in space limit the deviation away from the Cassini plane to a phase lag of approximately 10 arcsec, and our results show that this implies that the mantle viscosity cannot be much smaller than $10^{17}$ Pa s. 

\section{Introduction}

\begin{figure}
\begin{center}
    \includegraphics[height=6.5cm]{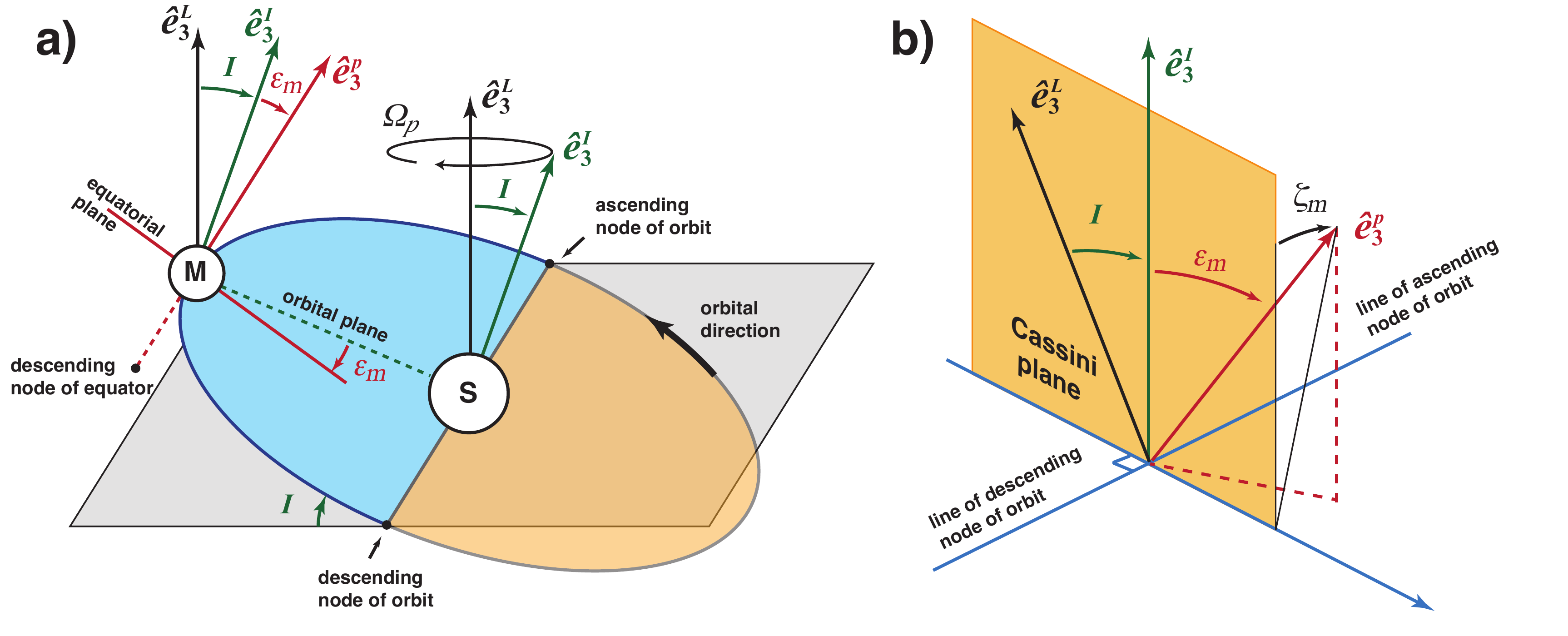} 
    \caption{\label{fig:cassini}  The Cassini state of Mercury.  (a) The orbit of Mercury (M) around Sun (S) with respect to the Laplace plane (grey shaded rectangle) and the Cassini state of Mercury.  The normal to the orbital plane ($\mitbf{\hat{e}_3^I}$) is offset from the normal to the Laplace plane ($\mitbf{\hat{e}_3^L}$) by an angle $I=8.5330^\circ$.  The symmetry axis of the mantle $\mitbf{\hat{e}_3^p}$ (assumed to be exactly aligned with the mantle rotation vector in this cartoon) is offset from $\mitbf{\hat{e}_3^I}$ by an obliquity angle of $\varepsilon_m\approx2$ arcmin.  Both $\mitbf{\hat{e}_3^I}$ and $\mitbf{\hat{e}_3^p}$ precess about $\mitbf{\hat{e}_3^L}$ in a retrograde direction at frequency $\Omega_p = 2\pi/325,513$ yr$^{-1}$.  The blue (orange) shaded region indicates the portion of the orbit when Mercury is above (below) the Laplace plane. (b) In an ideal Cassini state, $\mitbf{\hat{e}_3^p}$ lies in the plane defined by $\mitbf{\hat{e}_3^L}$ and $\mitbf{\hat{e}_3^I}$ (the Cassini plane, orange shaded rectangle).  Dissipation of rotational energy displaces $\mitbf{\hat{e}_3^p}$ out of the Cassini plane by a phase-lag angle $\zeta_m$.  In the complex notation used in our study, $\zeta_m = Im[\tilde{\varepsilon}_m]$.  Angles in both panels are not drawn to scale but exaggerated for the purpose of illustration.} 
\end{center}
\end{figure}

\begin{figure}
\begin{center}
    \includegraphics[height=7.1cm]{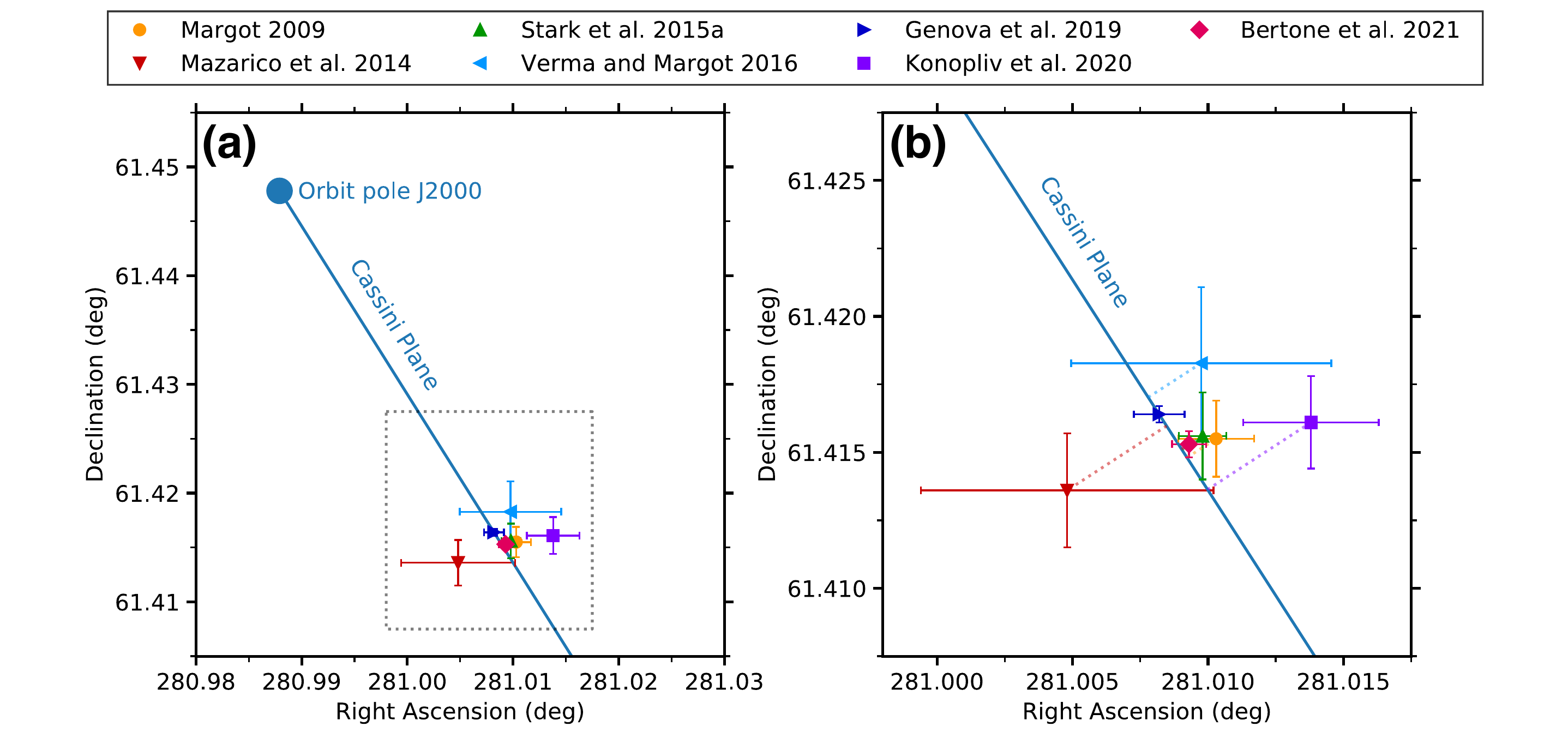}    
    \caption{\label{fig:cass2} a) Right ascension and declination angles of the spin pole of Mercury based on different studies. The location of the orbit pole and the orientation of the Cassini plane are taken from \citet{baland17}. b) Close up view near the spin pole locations. The horizontal and vertical lines indicate the 1$\sigma$ errors on right ascension and declination, respectively, and the dashed lines indicate the deviation to the Cassini plane.  This deviation is the phase lag angle $\zeta_m$, positive (negative) for spin poles measurements located to the top-right (bottom-left) of the Cassini plane line.} 
\end{center}
\end{figure}

The spin axis of Mercury is in a Cassini state (Figure \ref{fig:cassini}).  The latter describes a configuration in which the planet's spin axis and orbit normal remain coplanar to and precess about the normal to the Laplace plane \cite[][]{colombo66,peale69,peale06}.  The precession is retrograde, and the latest estimate of its period is $325,513 \pm 10,713$ years \cite[][]{baland17}.  Figure \ref{fig:cass2} shows the orientation in space of the spin axis reported in several recent studies, expressed at the J2000 epoch as is the usual convention.  A visual inspection of Figure \ref{fig:cass2} reveals that, within measurement errors, Mercury's spin axis aligns with the plane defined by the Laplace pole and orbit normal, a plane which we refer to as the Cassini plane, confirming that Mercury occupies a Cassini state.

The retrograde precession of the Cassini plane implies that the line that depicts its location in Figure \ref{fig:cass2} is displaced toward the bottom-left as a function of time. Hence, a spin pole located to the top-right (bottom-left) with respect to this line is behind (ahead of) the expected Cassini state orientation, and corresponds to a phase lag (phase lead).  We denote the offset from the Cassini plane by an angle $\zeta_m$, defined positive for a phase lag (see Figure \ref{fig:cassini}b). Table \ref{tab:lag} gives the spin pole orientations from the recent measurements that are plotted in Figure \ref{fig:cass2}, as well as their phase lag angles $\zeta_m$, calculated by the method described in Appendix A.   For all spin pole measurements, the 1$\sigma$ error on the phase lag is either larger than the phase lag itself, or of similar magnitude.  This confirms that, within measurement errors, Mercury's spin pole indeed occupies a Cassini state.  The magnitude of the phase lags in Table \ref{tab:lag} provides a quantitative measure of the deviation from an exact Cassini state. For all spin pole measurements, the phase lag is smaller than 10 arcsec; it is smaller than 1 arcsec for two of the most recent measurements (those of \citet{genova19} and \citet{bertone21}).  The only measurement that suggest a phase lead ($\zeta_m<0$) is that from the study of \citet{mazarico14}.

If the Mercury-Sun system were to be taken in isolation and if the dissipation of rotational and orbital energy were negligibly small, Mercury would obey an exact Cassini state.   However, in reality, small deviations from an exact Cassini state are expected.   First, the precession of Mercury's pericentre at a period of 134,477 yr induced by gravitational forces from other planets generates a small nutation motion of approximately 0.85 arcsec of the spin axis with respect to its position in the Cassini state \cite[][]{baland17}.  At epoch J2000, the phase of this nutation is such that  the spin axis is displaced approximately perpendicular to (out-of) the Cassini plane, towards the top-right quadrant of Figure \ref{fig:cass2}.  As a result, the spin axis should not be aligned exactly with the Cassini plane, but it should lag behind it by an angle of $\zeta_m \approx 0.85$ arcsec.  

An additional deviation from the Cassini plane is expected from the dissipation of orbital and rotational energy which, even if small, is invariably present.  Indeed, a Cassini state configuration is a state of minimum energy, and can only be attained as a result of dissipation of an earlier more energetic state.  One source of dissipation is from tidal deformations that occur in response to the solar gravitational potential imposed on Mercury.  Tidal deformations are never perfectly elastic, some of the energy being dissipated as heat within the planet.  Tidal dissipation is characterized by a quality factor $Q$.  As a reference, for $Q\approx 100$, a reasonable planetary value, tidal dissipation should induce a phase lag angle of $\zeta_m \approx 1$ arcsec \cite[][]{baland17}.  A smaller $Q$ would induce a larger $\zeta_m$ and conversely, a larger $Q$ would induce a smaller $\zeta_m$. 

Taken together, the deviation away from the Cassini plane induced by the precession of the pericentre and tidal dissipation (based on $Q\approx100$) should lead to a phase lag of $\zeta_m \approx 1.85$ arcsec.  This is approximately equal to the error in $\zeta_m$ derived from the spin pole orientation measurement of \citet{bertone21}.  If we take this latter measurement as a benchmark, this implies that $Q$ cannot be much smaller than 100.

Another source of dissipation is viscous and electromagnetic (EM) drag at the core-mantle boundary (CMB) and inner core boundary (ICB) of Mercury.  If the core of Mercury were fully solidified, the orientation of the spin (and symmetry) axis depicted in Figures \ref{fig:cassini} and \ref{fig:cass2} would characterize that of the entire planet.  However, the electrically conducting core must be partially fluid, as motions within it are required to sustain Mercury's internally generated magnetic field \cite[][]{anderson11,anderson12,johnson12}.  The observed amplitude of Mercury's 88-day libration provides additional support for a partially fluid core \cite[e.g.][]{margot07,margot12}.  Just like the Earth, the central region of Mercury's core may be solid, although the size of this solid inner core, if it exists, is not well constrained \cite[e.g.][]{steinbrugge21}.  The measurements shown in Figure \ref{fig:cass2} reflect then the orientation of the spin (and symmetry) axis of Mercury's outer solid shell comprised of its mantle and crust.  We do not have direct measurements of the orientation of the spin axis of the fluid core nor, if present, that of the solid inner core.  However, we expect that their spin axes also obey a Cassini state, though with different obliquity angles than that of the mantle \cite[e.g.][]{peale14,peale16,dumberry21}.  The differentially rotating mantle, fluid core and inner core imply viscous drag at the CMB and at the ICB.  The shearing of the magnetic field threading the ICB also leads to EM drag, a process that also occurs at the CMB if the lowermost region of the mantle is electrically conducting. 

Dissipation from viscous and EM drag at the CMB and ICB drains some of Mercury's rotational energy and, consequently, induces a deviation of the spin pole away from the Cassini plane.  The magnitude of these internal sources of dissipation, and hence the resulting $\zeta_m$, depend on parameters that are not well known, including the viscosity of the fluid core, the electrical conductivity of both the inner and fluid cores, and the strength of the magnetic field inside the core.  However, predictions can be computed based on a range of model parameters.  The calculations presented in \citet{peale14} suggest that viscous and/or EM coupling may amount to a phase lag of 0.05 arcsec.   Clearly, the total dissipation from the combined effects of tidal deformations and viscous and EM friction at the boundaries of the fluid core must be limited, as otherwise the spin pole would deviate from the Cassini plane by a greater angle than the upper bound of a few arcsec suggested by measurements.

The main objective in this work is to compute estimates of the dissipation and phase lag angle $\zeta_m$ induced by tidal deformation and by viscous and EM drag at the CMB and ICB of Mercury.  A model to compute the Cassini state of Mercury comprising a fluid core and solid inner core is presented in detail in \cite{dumberry21} (referred to hereafter by D21).  This model includes viscous and EM coupling at the ICB and CMB. The focus in D21 was on the effects that viscous and EM coupling have on the obliquity angle, in other words on the component of the spin pole orientation contained in the Cassini plane.  The present work can be thought of as the second part of D21, focused here on the component of the spin pole out of the Cassini plane.  We provide an update on the predictions of $\zeta_m$ made by \cite{peale14} due to viscous drag (which dominates EM drag) at the CMB and complemented by the inclusion of EM drag (which dominates viscous drag) at the ICB.  The model developed in D21 did not include viscoelastic deformations induced by tidal forces and by the differential rotation of Mercury's interior regions.  We modify here the model in D21 to include these effects.  A connection between $\zeta_m$ and the tidal quality factor $Q$ is presented in \citet{baland17}; our model is consistent with their results, and we make an additional effort to relate $Q$ to the viscosities of the mantle and inner core.  

As Table \ref{tab:lag} illustrates, current measurements of the spin pole orientation are not sufficiently precise to determine the phase lag with high accuracy.  Hence, we do not have a specific observational target that we aim to match.  Our study is instead an exploration of the different dissipative mechanisms and the phase lag they produce.  It is likely that the accuracy of the spin pole orientation will improve with future observations, such as that from the upcoming BepiColombo satellite mission \cite[][]{cicalo16}. Predictions of the phase lag by a combination of tidal dissipation and viscous and EM drag at the CMB and ICB may provide an opportunity to further constrain the internal dissipation taking place within Mercury, and in turn, the physical parameters associated with these processes.

\begin{table}
\begin{tabular}{llll}
\hline
Study & Right ascension ($^\circ$) & Declination ($^\circ$) & Phase lag (arcsec)  \\ \hline
\cite{margot12} &  $281.0103 \pm 0.0015$ & $61.4155 \pm 0.0013$ & $2.50 \pm 2.83$ \\
\cite{mazarico14} &  $281.0048 \pm 0.0054$ & $61.41436 \pm 0.0021$ & $-7.76 \pm 9.16$  \\
\cite{stark15} &  $281.00980 \pm 0.00088$ & $61.4156 \pm 0.0016$ & $1.79 \pm 2.23$  \\
\cite{verma16} &  $281.00975 \pm 0.0048$ & $61.41828 \pm 0.0028$ & $4.55 \pm 8.44$  \\
\cite{genova19} &  $281.0082 \pm 0.0009$ & $61.4164 \pm 0.0003$ & $0.00364 \pm 1.52$  \\
\cite{konopliv20} &  $281.0138 \pm 0.0025$ & $61.4161 \pm 0.0017$ & $8.90 \pm 4.49$  \\
\cite{bertone21} &  $281.0093 \pm 0.00063$ & $61.4153 \pm 0.00048$ & $0.645 \pm 1.15$  \\
\hline
\end{tabular}
\caption{\label{tab:lag} Right ascension, declination and phase lag angle with respect to the Cassini plane at J2000 for recent measurements of Mercury's spin pole orientation. The phase lag angles give the distance to the Cassini plane of the central value of each of the spin pole orientation measurements.  See Appendix A for details of the calculations of the phase lags and their estimated errors. }
\end{table}


\section{Theory} 

The rotational model of Mercury that we use and the way we construct interior models of Mercury are presented in detail in D21.  For convenience we briefly mention some of their salient features below.  We modify the rotational model of D21 to take into account viscoelastic deformations.  These modifications are presented in Appendix B. 

\subsection{Interior structure}

Mercury (mass $M$) is modelled as a simple four layer planet comprised of an inner core, fluid core, mantle, and crust, each with a uniform density.   The outer spherical mean radii of each of these layers, are denoted by $r_s$, $r_f$, $r_m$, and $R$, and their densities by $\rho_s$, $\rho_f$, $\rho_m$, and $\rho_c$, respectively. The inner core radius $r_s$ corresponds to the ICB radius, the fluid core radius $r_f$ to the CMB radius, and $R$ to the planetary radius of Mercury. 

For the crust, we assume a density of $\rho_c=2974$ kg m$^{-3}$ and a thickness of $h=R-r_m = 26$ km \cite[][]{sori18}.  Individual interior models are constructed for each choice of ICB radius, ensuring that they are consistent with $M$ and chosen values of the moments of inertia of the whole planet $C$ and that of the combined mantle and crust $C_m$.  The latter two are determined from the observed obliquity $\varepsilon_m$ and the observed amplitude of the 88-day longitudinal librations.  We use here the same choices of $C$ and $C_m$ as in D21: $C/ MR^2 = 0.3455$ and $C_m/ MR^2 = 0.1475$.  Two possible end-member scenarios for how the densities of the solid ($\rho_s$) and fluid ($\rho_f$) cores may evolve with inner core growth were considered in D21.  In the first, $\rho_s$ is held constant and $\rho_f$ is adjusted with inner core size to match $M$.  This captures a Fe-S core composition with little or no S being incorporated into the inner core as it crystallizes.  In the second scenario, it is the density contrast at the ICB which is set to a constant, capturing a Fe-Si core composition in which Si is expected to partition into the solid core.  Specific solutions of the rotational model depend on which of these scenarios is used, but their qualitative behaviour are equivalent.  Numerical results are computed here according to the first scenario, with $\rho_s=8,800$ kg m$^{-3}$.

Each layer is triaxial in shape.  We define the polar geometrical ellipticity of each layer as the difference between the mean equatorial and polar radii, divided by the mean spherical radius.  Likewise, we define the equatorial geometrical ellipticity of each layer as the difference between the maximum and minimum equatorial radii, divided by the mean spherical radius.  The polar and equatorial geometrical ellipicities are denoted by $\epsilon_i$ and $\xi_i$ respectively, with the subscript $i$ = $s$, $f$, $m$, and $r$ denoting the ICB, CMB, crust-mantle boundary, and surface, respectively.  The polar and equatorial flattenings at the surface are taken from \cite{perry15} and their values are given in Table 1 of D21.  We assume that the shapes of the ICB and CMB coincide with equipotential surfaces at hydrostatic equilibrium, and the flattenings at all interior boundaries are specified such that they match the observed degree 2 spherical harmonic coefficients of gravity $J_2$ and $C_{22}$ (their numerical values are given in Table 1 of D21).

With the densities and ellipticities of each interior regions known, one can compute the moments of inertia of the fluid core ($C_f > B_f > A_f$) and solid inner core ($C_s > B_s > A_s$).  The rotational model involves the mean equatorial moments of inertia $\bar{A}, \bar{A}_f, \bar{A}_s$ of the whole planet, fluid core and solid inner core and the dynamical ellipticities $e$, $e_f$, $e_s$, $\gamma$ and $\gamma_s$.  These are defined and computed according to Equations 2 and 3 of D21.

\subsection{Rotational model}

Mercury rotates in a 3:2 spin-orbit resonance.  Its sidereal frequency $\Omega_o=2\pi/58.64623$ day$^{-1}$ is 1.5 times its orbital frequency (or, mean motion) $n=2\pi/87.96935$ day$^{-1}$ \cite[][]{stark15b}.   Mercury's rotation is also characterized by a Cassini state.  The latter defines a configuration in which the orientations of the normal to the orbital plane (or, orbital pole, $\mitbf{\hat{e}_3^I}$) and the symmetry axis ($\mitbf{\hat{e}_3^p}$) are both coplanar with, and precess about, the normal to the Laplace plane (or, Laplace pole, $\mitbf{\hat{e}_3^L}$).  The rotation vector of Mercury $\boldsymbol{\Omega}$ is not exactly aligned with the symmetry axis $\mitbf{\hat{e}_3^p}$  in the Cassini state equilibrium, but the offset between the two is small, approximately 0.015 arcsec (see Equation \ref{eq:mepsm} below). The Cassini state of Mercury is illustrated in Figure \ref{fig:cassini}. The orientation of the Laplace pole varies on long timescales, but it is convenient here to assume that it is invariant in inertial space.  The precession of $\mitbf{\hat{e}_3^I}$ and $\mitbf{\hat{e}_3^p}$ about the Laplace normal is retrograde with frequency $\Omega_p = 2\pi/325,513$ yr$^{-1}$ \cite[][]{baland17}. 

Since Mercury has a fluid core and (possibly) a solid inner core, $\mitbf{\hat{e}_3^p}$ and $\boldsymbol{\Omega}$ characterize the symmetry and rotation axes of the solid shell of Mercury comprised of its mantle and crust. Three additional orientation vectors are required to fully describe the Cassini state: the rotation vectors of the fluid core ($\boldsymbol{\Omega_f}$) and inner core ($\boldsymbol{\Omega_s}$) and the symmetry axis of the inner core ($\mitbf{\hat{e}_3^s}$) (see Figure 2 of D21); these also precess in the retrograde direction with frequency $\Omega_p$ about the Laplace pole.

The specific orientation of each of the vectors $\mitbf{\hat{e}_3^p}$, $\boldsymbol{\Omega}$, $\boldsymbol{\Omega_f}$, $\boldsymbol{\Omega_s}$ and $\mitbf{\hat{e}_3^s}$ in the Cassini state equilibrium depends on the mean solar torque (time-averaged over one orbit) applied on Mercury's instantaneous figure and on internal torques that arise from the misalignment between its interior regions. The rotational model in D21 solves for these orientations. It consists of a linear system of five equations written in terms of five rotational variables, $\tilde{\varepsilon}_m$, $\tilde{m}$, $\tilde{m}_f$, $\tilde{m}_s$ and $\tilde{n}_s$, which are projections of the five orientation vectors in the equatorial plane of Mercury's rotating frame.  

In the absence of dissipation, the vectors $\mitbf{\hat{e}_3^p}$, $\boldsymbol{\Omega}$, $\boldsymbol{\Omega_f}$, $\boldsymbol{\Omega_s}$ and $\mitbf{\hat{e}_3^s}$ all lie in the Cassini plane.  Viewed in the inertial frame, the Cassini plane is rotating in a retrograde direction at frequency $\Omega_p$.  The equations of the rotational model of D21 are developed in a frame attached to the mantle and crust rotating at sidereal frequency $\Omega_o$. Viewed in this frame, the Cassini plane is rotating in a retrograde direction at frequency $\omega \Omega_o$ (see Figure 2b of D21), where $\omega$, expressed in cycles per Mercury day, is equal to (Equation 21 of D21)

\begin{equation}
\omega = -1 - \delta \omega \cos I \, ,
\label{eq:omega1}
\end{equation}
where $I=8.5330^\circ$ is the inclination of the orbital plane.  The factor $\delta \omega = \Omega_p/\Omega_o = 4.933 \times 10^{-7}$ is the Poincar\'e number, the ratio of the forced precession to sidereal rotation frequencies.  The mean solar torque is pointing in the same direction as the vector connecting the Sun to the descending node of Mercury's orbit (see Figure \ref{fig:cassini}), so from the mantle-fixed frame the orientation of this mean torque is periodic, rotating at frequency $\omega \Omega_o$.  Setting the equatorial directions $\mitbf{\hat{e}_1^p}$ and $\mitbf{\hat{e}_2^p}$ to correspond with the real and imaginary axes of the complex plane, respectively, the equatorial components of the mean solar torque is written in a compact form as

\begin{equation}
{\Gamma}_1(t) + i {\Gamma}_2(t) = - i  \, \tilde{\Gamma}(\omega) \,  \exp[{i \omega \Omega_o t}] \, , \label{eq:gammaphi}
\end{equation}
where $i=\sqrt{-1}$ and $\tilde{\Gamma}(\omega)$ represents the amplitude of the torque at frequency $\omega \Omega_o$.  The rotational variables $\tilde{\varepsilon}_m$, $\tilde{m}$, $\tilde{m}_f$, $\tilde{m}_s$ and $\tilde{n}_s$ are complex amplitudes, also proportional to $\exp[{i \omega \Omega_o t}]$, in response to this applied external torque.  Their real parts correspond to the angles of the five rotational vectors in the Cassini plane (i.e. in-plane components), the response that is in-phase with the applied solar torque.  Their imaginary parts reflect the component of these angles out of the Cassini plane (out-of-plane components), the out-of-phase response to the applied torque as a result of dissipation. A positive imaginary part corresponds to a phase lag, a negative imaginary part to phase lead.  

The rotational model of D21 includes a parameterization for the viscous and EM torques at the CMB and ICB expressed as 

\begin{subequations}
\begin{equation}
    \tilde{\Gamma}_{cmb} = i \Omega_o^2 \bar{A}_f K_{cmb} \, \tilde{m}_f \, ,\label{eq:tqcmb}
\end{equation}
\begin{equation}
    \tilde{\Gamma}_{icb} = i \Omega_o^2 \bar{A}_s K_{icb} (\tilde{m}_f - \tilde{m}_s) \, ,\label{eq:tqicb}
\end{equation}
\label{eq:tqvisc}
\end{subequations}
where $K_{cmb}$ and $K_{icb}$ are dimensionless complex coupling constants. Specific expressions for the viscous and EM coupling models are given further ahead in the results sections.  These torques generate both an in-phase and out-of-phase response.  

The model of D21 assumes a rigid outer shell (mantle and crust) and a rigid inner core.  Here, we take into account viscoelastic deformations within each interior region in response to gravitational and centrifugal forces.  Such deformations induce a perturbation in the moment of inertia tensors of each region and therefore a modification of both the solar torque and Mercury's angular momentum response.  The details of how the rotational model is adapted to include these are presented in Appendix B.  Deformations are characterized by a set of compliances ${\cal S}_{ij}$ which quantify the changes in the moment of inertia tensors of each region.  

Elastic tidal deformations of a planetary body are typically expressed by the Love number $k_2$.  The latter represents the fractional change in the gravitational potential of degree 2 at the surface induced by global  deformations. Viscous or anelastic deformations are captured by a quality factor $Q$, with $Q^{-1}$ representing the fraction of the total energy that is dissipated over one cycle.  A low (high) Q value indicates a high (low) dissipation. $k_2$ and $Q^{-1}$ characterize, respectively, deformations that are in-phase and out-of-phase with the tidal potential. In our rotational model, these are connected to the compliance ${\cal S}_{11}$ through 

\begin{equation}
    Re[{\cal S}_{11}] = k_2 \frac{R^5\Omega_0^2}{3G\bar{A}} \, , \hspace{1cm}     Im[{\cal S}_{11}] = \frac{k_2}{Q} \frac{R^5\Omega_0^2}{3G\bar{A}} \, ,\label{eq:k2Q}
\end{equation}
where $G$ is the gravitational constant.   Recent estimates of $k_2$ are $0.569\pm0.025$ \cite[][]{genova19} and $0.53\pm0.03$ \cite[][]{konopliv20}.  We do not have direct observational constraints on $Q$.  

The method to compute the compliances ${\cal S}_{ij}$ is presented in Appendix C.  Their numerical values depend on the rheology assumed in the solid regions (crust, mantle and inner core).  We assume a Maxwell solid rheology, and constrain this rheology such that $k_2$ in all our interior models matches $k_2 = 0.55$, a value at the mid-point of the recent estimates given above. The quality factor $Q$ depends on the uniform viscosity assumed within the mantle and inner core; we present results for a range of possible values.  To give a sense of the amplitude of $S_{11}$, we can approximate $\bar{A}$ to be equal to the mean (spherical) moment of inertia and take the latter to be $0.346\cdot MR^2$ \cite[][]{margot12}.  Using the parameters from Table 1 of D21, a tidal Love number $k_2 = 0.55$ (the value that we use for all our results), corresponds to $Re[{\cal S}_{11}] = 5.37 \times 10^{-7}$.  For $Q=100$, this gives $Im[{\cal S}_{11}] = 5.37 \times 10^{-9}$.

\subsection{Approximate solutions}

The set of equations that enter the rotational model is presented in Appendix B.  Substituting $\omega = -1 - \delta \omega \cos I$ (Eq. \ref{eq:omega1}) in Equations (\ref{eq:kinI}) and (\ref{eq:msns}) provides the following two kinematic relationships, relating $\tilde{m}$ to $\tilde{\varepsilon}_m$ and $\tilde{m}_s$ to $\tilde{n}_s$: 

\begin{subequations}
\begin{align}
\tilde{m} & = \delta \omega (\sin I + \tilde{\varepsilon}_m \cos I) \, , \label{eq:mepsm}\\
\tilde{m}_s & = (1+\delta \omega \cos I) \tilde{n}_s \, .
\end{align}
\end{subequations}
With $I=8.5330^\circ$, $\delta \omega = 4.9327 \times 10^{-7}$ and taking $\tilde{\varepsilon}_m=2.04$ arcmin, this gives $\tilde{m}= 0.0151$ arcsec: the offset of the spin axis of the mantle with respect to its symmetry axis is very small.  Similarly, the misalignment between the spin axis of the inner core ($\tilde{m}_s$) and its symmetry axis ($\tilde{n}_s$) is also very small: as an indication, for an inner core tilt with respect to the mantle of $\tilde{n}_s=1$ arcmin, $\tilde{m}_s$ is offset from $\tilde{n}_s$ by approximately 0.03 milliarcsec.  

For the purpose of building an approximate analytical solution, we can simply assume $\tilde{m}_s=\tilde{n}_s$.  However, we cannot set $\tilde{m}=0$.  This is because our system of equations is developed in the frame of the rotating mantle.  In this frame, $\tilde{m}$ captures the change in mantle angular momentum induced by the solar torque.  To express this change in terms the orientation of Mercury's figure in the inertial (Laplace) frame, we substitute $\tilde{m}$ with Equation \ref{eq:mepsm}.

Approximate solutions for the obliquity and phase lag of the mantle can be constructed from the angular momentum equation for the whole of Mercury (Equation \ref{eq:am1}). All compliances ${\cal S}_{ij}$ are of the order of $10^{-7}$ or smaller; the term $\tilde{c}/\bar{A}$ can be neglected when compared to other terms on the left-hand side.  By substituting Eq. \ref{eq:mepsm} and setting $\tilde{m}_s=\tilde{n}_s$, we can simplify Eq. (\ref{eq:am1}) to

\begin{equation}
 -\frac{C}{\bar{A}}\,  \delta \omega \Big( \sin I + \tilde{\varepsilon}_m \cos I \Big)  -\delta \omega \cos I \Bigg[ \frac{\bar{A}_f}{\bar{A}} \tilde{m}_f + \frac{\bar{A}_s}{\bar{A}} \tilde{n}_s \Bigg] =    \frac{1}{i \Omega_o^2 \bar{A}} \Big(\tilde{\Gamma}_{sun} +\tilde{\Gamma}_{t}  \Big) \, , \label{eq:am1b}
 \end{equation}
 where we have used $C =\bar{A} (1+e)$, and where the torques $\tilde{\Gamma}_{sun}$ and $\tilde{\Gamma}_{t}$ are given by Equations (\ref{eq:tqsun6}) and (\ref{eq:td3}).  Keeping only the largest terms in the former, these are given by

\begin{subequations}
\begin{align}
\frac{\tilde{\Gamma}_{sun}}{i \Omega_o^2 \bar{A}} & = -\left[ \phi_m^{el}\,  \tilde{\varepsilon}_m + \frac{\bar{A}_s}{\bar{A}} \alpha_3 \phi_s^{el} \tilde{n}_s   + \frac{\phi_m}{e} \Big( {\cal S}_{12} \tilde{m}_f + {\cal S}_{14} \tilde{n}_s \Big) \right]  \, ,\label{eq:tqsunmain} \\
\frac{\tilde{\Gamma}_t}{i \Omega_o^2 \bar{A}} & =   \, i Im[{\cal S}_{11}] \Big[ \phi_m^{t3} \tilde{\varepsilon}_m  +   \phi_m^{t2} \cos I \sin I \Big] \, , \label{eq:tdmain}
\end{align}
\end{subequations}
where $\alpha_3 = 1 -\rho_f/\rho_s$ is the density contrast at the ICB.  The definitions of the torque factors $\phi_m$, $\phi_m^{el}$, $\phi_s^{el}$, $\phi_m^{t2}$ and $\phi_m^{t3}$ are given in Appendix B.  In addition to the compliance ${\cal S}_{11}$, the two additional compliances that have the largest influence on the solutions are ${\cal S}_{12}$ and ${\cal S}_{14}$. These capture the global viscoelastic deformations of Mercury in response to internal forcing.  For ${\cal S}_{12}$, it is the centrifugal force on the CMB by the misaligned spin axis of the fluid core.  For ${\cal S}_{14}$, it is the gravitational force from the tilted inner core.  The compliances are complex: their real and imaginary parts capture, respectively, elastic and anelastic deformations.   

Using $\delta \omega = \Omega_p/\Omega_o$, with Equations (\ref{eq:tqsunmain}-\ref{eq:tdmain}), Equation (\ref{eq:am1b}) can be written as

\begin{align}
C \Omega_p \Big( \sin I + \tilde{\varepsilon}_m \cos I \Big)  + \Omega_p \cos I \Big( \bar{A}_f \tilde{m}_f + \bar{A}_f \tilde{m}_f \Big) & = \nonumber\\ 
 & \hspace*{-3cm} \bar{A} \Omega_o \phi_m^{el} \,\tilde{\varepsilon}_m + \bar{A}_s \Omega_o \alpha_3 \phi_s^{el} \, \tilde{n}_s + \bar{A} \Omega_o\frac{\phi_m}{e} \Big( {\cal S}_{12} \tilde{m}_f + {\cal S}_{14} \tilde{n}_s \Big)  \nonumber\\
 & \hspace*{-3.4cm} - i \bar{A} \Omega_o Im[{\cal S}_{11}] \bigg( \phi_m^{t3} \tilde{\varepsilon}_m + \phi_m^{t2} \sin I \cos I \bigg) \, . \label{eq:approxsol}
\end{align}
From this latter equation, we can derive approximate solutions for both the obliquity (in-plane component) $\varepsilon_m=Re[\tilde{\varepsilon}_m]$ and the phase lag (out-of-plane component) $\zeta_m = Im[\tilde{\varepsilon}]$.  

\subsection{Obliquity}

Although our study focuses on the phase lag, the introduction of viscoelastic deformations in the rotational model alters the obliquity solutions presented in D21.  For completeness, let us first consider predictions of the obliquity, which can be computed from the real part of Equation (\ref{eq:approxsol}), and can be written as 

\begin{equation}
    \varepsilon_m = \varepsilon_m^{t} + \varepsilon_m^{L,c} + \varepsilon_m^{t,s} + \varepsilon_m^{t,e} + \varepsilon_m^{t,a} \, , \label{eq:eps1}
\end{equation}
where

\begin{subequations}
\begin{align}
\varepsilon_m^{t} &= \frac{C \Omega_p \sin I}{{\cal L}_m}  \, , \\ 
\varepsilon_m^{L,c} &= \frac{\bar{A} \Omega_p \cos I}{{\cal L}_m} \bigg[ \frac{\bar{A}_f}{\bar{A}} Re[\tilde{m}_f]  + \frac{\bar{A}_s}{\bar{A}} Re[\tilde{n}_s]   \bigg] \, , \\ 
\varepsilon_m^{t,s} &= \frac{\bar{A}_s \Omega_o}{{\cal L}_m} \bigg[- \alpha_3 \phi_s^{el} Re[\tilde{n}_s] \bigg] \, , \\ 
\varepsilon_m^{t,e} &= \frac{\bar{A} \Omega_o}{{\cal L}_m} \frac{\phi_m}{e} \bigg[ - Re[S_{12}] Re[\tilde{m}_f]  - Re[S_{14}] Re[\tilde{n}_s]  \bigg]  \,  ,\\ 
\varepsilon_m^{t,a} &= \frac{\bar{A} \Omega_o}{{\cal L}_m} \frac{\phi_m}{e} \bigg[  Im[S_{12}] Im[\tilde{m}_f]  + Im[S_{14}] Im[\tilde{n}_s]  \bigg] \, ,
\end{align}
\label{eq:varepsmmulti}
and 

\begin{equation}
{\cal L}_m  = \bar{A} \Omega_o  \phi_m^{el} - C \Omega_p \cos I \, .
\end{equation}
\end{subequations}
Each of the terms on the right-hand side of Equation (\ref{eq:eps1}) captures a contribution to $\varepsilon_m$ from a different origin.   $\varepsilon_m^{t}$ captures the obliquity resulting from the solar torque acting on the ellipsoidal shape of Mercury.  $\varepsilon_m^{L,c}$ captures the contribution to the obliquity connected with the angular momentum carried by the fluid and solid cores.  These result from internal torques between the mantle, fluid core and solid core; this term captures then the mantle obliquity generated by internal torques.  The remaining three contributions result from the solar torque acting on additional aspherical features of Mercury's shape. In $\varepsilon_m^{t,s}$, it is on the tilt of the ellipsoidal figure of the inner core with respect to the mantle. In $\varepsilon_m^{t,e}$, it is on the global elastic deformation caused by the in-plane components of the misaligned fluid core spin axis ($\tilde{m}_f$) and inner core tilt ($\tilde{n}_s$).  In  $\varepsilon_m^{t,a}$, it is on the delayed, anelastic deformation in response to the out-of-plane components of $\tilde{m}_f$ and $\tilde{n}_s$.   

In the absence of a fluid core and inner core, 

\begin{equation}
    \varepsilon_m = \varepsilon_m^t = \frac{C \Omega_p \sin I }{{\cal L}_m } = \frac{C \Omega_p \sin I }{ \bar{A} \Omega_o  \phi_m^{el} - C \Omega_p \cos I} \, . 
\end{equation}
This is identical to Equation (26) of D21, except that $\phi_m$ has been replaced by $\phi_m^{el}$; the latter is a modification of the former by elastic deformation (see Equation \ref{eq:phimel}).  We also retrieve, in our notation, the solution given in Equation (64) of \citet{baland17}, where their definition of $\dot{\Omega}$ is equal to $-\Omega_p$. [Note also that their definition of $\tilde{C}$ is equal, in our notation, to $C - \bar{A} (\phi_m/e) Re[{\cal S}_{11}]$, which differs from $C$ only by a few parts in $10^{7}$ and can be neglected.]

The real and imaginary parts of $\tilde{m}_f$ and $\tilde{n}_s$ can be similar in magnitude for sufficiently strong viscous or EM coupling at the ICB and CMB.  However, the imaginary parts of the compliances are smaller than their real parts by a factor approximately equal to the quality factor $Q$.  Hence, provided that $Q>10$, this implies that $\varepsilon_m^{t,e}\gg\varepsilon_m^{t,a}$.  For a small or no inner core, $\bar{A}_s \ll \bar{A}_f$, ${\cal S}_{14} \ll {\cal S}_{12}$ and the prediction of the obliquity is

\begin{equation}
    \varepsilon_m = \frac{C \Omega_p \sin I }{{\cal L}_m }  + \frac{\bar{A} \Omega_o}{{\cal L}_m} \left( \frac{\bar{A}_f}{\bar{A}} \frac{\Omega_p}{\Omega_o} \cos I - \frac{\phi_m}{e} Re[{\cal S}_{12} ]\right) Re[\tilde{m}_f] \, . \label{eq:oblpredmf}
\end{equation} 
The second term on the right-hand side, connected to the misaligned spin axis of the core, is comprised of two parts with opposite signs; an angular momentum part, and a global deformation part.  Both $\bar{A}_f/\bar{A}$ and $\phi_m/e$ are fractions smaller than 1 (and of order 1), and the Poincar\'e number ($\delta \omega = \Omega_p /\Omega_o = 4.93 \times 10^{-7}$) is of the same order as $Re[{\cal S}_{12}]$ which is approximately equal to $3.5 \times 10^{-7}$.  Since $\bar{A}_f/\bar{A}<\phi_m/e$, not only is the term related to $Re[{\cal S}_{12}]$ non-negligible, it is larger in magnitude than the angular momentum part, and changes the sign of the correction to $\varepsilon_m$ associated with the misaligned spin axis of the fluid core. 

This is also true for the correction to $\varepsilon_m$ associated with the misaligned inner core: the part related to $Re[{\cal S}_{14}]$ is larger than the part related to its angular momentum.  In the contributions to $\varepsilon_m$, we thus have that $\varepsilon_m^{t,e}>\varepsilon_m^{L,c}$. Elastic deformations induced by the misaligned fluid core and solid core have to be taken into account in order to properly predict Mercury's obliquity.  

\begin{figure}
\begin{center}
 \includegraphics[height=6cm]{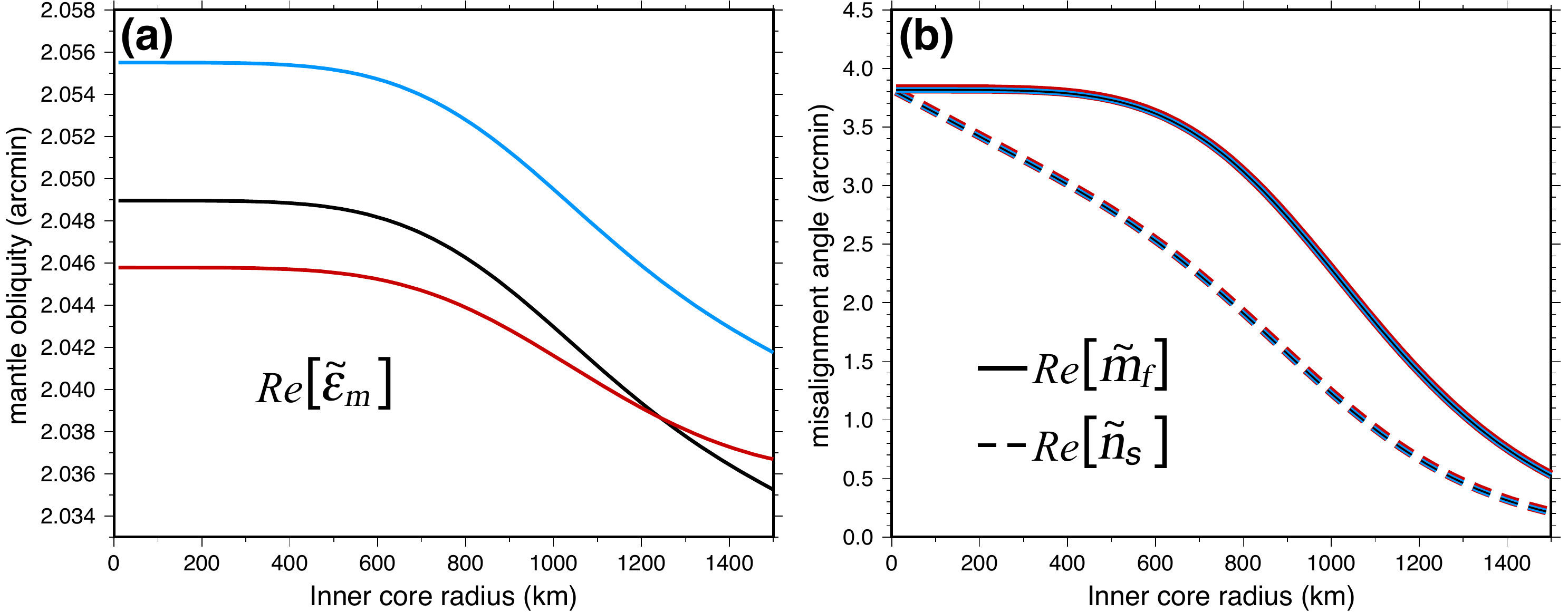} 
\caption{\label{fig:epsm} (a) Mantle obliquity ($Re[{\tilde{\varepsilon}}_m]$) and (b) misalignment angles of the fluid core spin axis ($Re[\tilde{m}_f]$, solid lines) and inner core figure axis ($Re[\tilde{n}_s]$, dashed lines) in the Cassini plane as a function of inner core radius.  Different colored lines correspond to solutions when all compliances ${\cal S}_{ij}$ are set to zero (black), when only ${\cal S}_{11}$ is retained (light blue), and when all compliances are included (red).  The thicknesses of the red and light blues curves have been increased in panel (b) to show that the different solutions of $Re[\tilde{m}_f]$ and $Re[\tilde{n}_s]$ are indistinguishable from one another on the scale of the figure.}
\end{center}
\end{figure}

We can illustrate for a specific example how the solutions presented in D21 are affected by the inclusion of the compliances ${\cal S}_{11}$, ${\cal S}_{12}$ and ${\cal S}_{14}$. Figure \ref{fig:epsm} shows how the real parts of $\tilde{\varepsilon}_m$, $\tilde{m}_f$ and $\tilde{n}_s$  vary with inner core size.  These solutions are computed with a viscosity in all solid regions equal to $10^{20}$ Pa s (i.e. in the elastic limit), $k_2=0.55$, a turbulent kinematic viscosity of $\nu = 10^{-4}$ m$^2$ s$^{-1}$ at both the ICB and CMB, an electrically insulating lowermost mantle (so that EM coupling at the CMB vanishes), an electrical conductivity of $10^{6}$ S m$^{-1}$ in both the solid and fluid cores, and a magnetic field strength at the ICB of $\left<B_r\right> = 0.1$ mT.  Three solutions are shown in Figure \ref{fig:epsm}.  First, a solution where all compliances ${\cal S}_{ij}$ are set to zero (black lines); the rotational model in this case is equivalent to that used in D21 and corresponds to a case where the crust, mantle and inner core are rigid.  Second, a solution where only the compliance ${\cal S}_{11}$ is retained (light blue lines).  Third, a solution that includes all compliances (red lines).  

Compared with the rigid case, the mantle obliquity $Re[\tilde{\varepsilon}_m]$ is increased by 0.0065 arcmin = 0.39 arcsec when the compliance ${\cal S}_{11}$ is introduced, reflecting the change in $Re[\tilde{\varepsilon}_m]$ caused by tidal deformations.  This is consistent with the results presented in \cite{baland17} (see their Figure 7), who also considered how tidal deformations (through the Love number $k_2$) affect the obliquity.  With the addition of all other compliances, compared to the solution when only ${\cal S}_{11}$ is retained, the mantle obliquity is reduced by 0.01 arcmin (for a small inner core) to 0.005 arcmin (for a large inner core).  It is dominantly the compliances ${\cal S}_{11}$, ${\cal S}_{12}$ and ${\cal S}_{14}$ that have an effect on the resulting mantle obliquity (the difference in the solution is virtually unchanged if only these three compliances are kept).  This third solution shows that elastic deformations induced by the misaligned spin axis of the fluid core (through ${\cal S}_{12}$) and the misaligned figure axis of the inner core (through ${\cal S}_{14}$) are as important as those from tidal forces on the resulting mantle obliquity. Present-day observations are not sufficiently precise to differentiate between the different solutions shown in Figure \ref{fig:epsm}a.  In other words, the observed mantle obliquity cannot be used to further constrain Mercury's rheology.  But if precision improves, our results illustrate that to do so properly, incorporating deformations caused by the misaligned fluid core and inner core in rotational models of the Mercury is necessary.  Finally, we note that the solutions of $\tilde{m}_f$ and $\tilde{n}_s$ (Figure \ref{fig:epsm}b) for these three different cases are virtually indistinguishable from one another; solutions of $\tilde{m}_f$ and $\tilde{n}_s$ for a rigid planet are not substantially different from those for a deformable planet.

\subsection{Phase lag}

The imaginary part of Eq. (\ref{eq:approxsol}) gives an approximate solution for the phase lag $\zeta_m = Im[\tilde{\varepsilon}_m]$, which can be written in a similar form as for the obliquity prediction, 

\begin{equation}
    \zeta_m = \zeta_m^{t} + \zeta_m^{L,c} + \zeta_m^{t,s} + \zeta_m^{t,a} + \zeta_m^{t,e} \, , \label{eq:zetam1}
\end{equation}
where

\begin{subequations}
\begin{align}
\zeta_m^{t} &= \frac{\bar{A} \Omega_o}{{\cal L}_m} Im[{\cal S}_{11}] \bigg[ \phi_m^{t3} Re[\tilde{\varepsilon}_m] + \phi_m^{t2} \sin I \cos I  \bigg] \, , \\ 
\zeta_m^{L,c} &= \frac{\bar{A} \Omega_p \cos I}{{\cal L}_m} \bigg[ \frac{\bar{A}_f}{\bar{A}} Im[\tilde{m}_f]  + \frac{\bar{A}_s}{\bar{A}} Im[\tilde{n}_s]   \bigg] \, , \\ 
\zeta_m^{t,s} &= \frac{\bar{A}_s \Omega_o}{{\cal L}_m} \bigg[- \alpha_3 \phi_s^{el} Im[\tilde{n}_s] \bigg] \, , \\ 
\zeta_m^{t,a} &= \frac{\bar{A} \Omega_o}{{\cal L}_m} \frac{\phi_m}{e} \bigg[ - Im[S_{12}] Re[\tilde{m}_f]  - Im[S_{14}] Re[\tilde{n}_s]  \bigg]  \, , \\ 
\zeta_m^{t,e} &= \frac{\bar{A} \Omega_o}{{\cal L}_m} \frac{\phi_m}{e} \bigg[ - Re[S_{12}] Im[\tilde{m}_f]  -  Re[S_{14}] Im[\tilde{n}_s]  \bigg] \, .
\end{align}
\label{eq:zetammulti}
\end{subequations}
The different contributions to $\zeta_m$ have similar physical interpretations to their counterparts for $\varepsilon_m$.   $\zeta_m^{t}$, $\zeta_m^{t,s}$, $\zeta_m^{t,e}$ and $\zeta_m^{t,a}$ capture the contributions to the phase lag from the solar torque acting on different out-of-plane aspherical features of Mercury.  In $\zeta_m^{t}$, it is on the delayed, anelastic tidal bulge of Mercury in response to the external gravitational force from the Sun. In $\zeta_m^{t,s}$, it is on the out-of-plane tilt of the inner core. In $\zeta_m^{t,a}$, it is on the delayed, anelastic deformation in response to the in-plane components of $\tilde{m}_f$ and $\tilde{n}_s$.  In $\zeta_m^{t,e}$, it is on the elastic deformation in response to out-of-plane components of $\tilde{m}_f$ and $\tilde{n}_s$.  $\zeta_m^{L,c}$ captures the contribution to the phase lag connected with the out-of-plane angular momentum carried by the fluid and solid cores.  

If we set $\tilde{m}_f=\tilde{n}_s=0$, which amounts to neglecting all contributions associated with the misaligned fluid core and solid inner core, the only contribution to the phase lag is from $\zeta_m^{t}$, and so

\begin{equation}
    \zeta_m = \zeta_m^{t} = \frac{\bar{A} \Omega_o}{{\cal L}_m} Im[{\cal S}_{11}] \bigg[ \phi_m^{t3} Re[\tilde{\varepsilon}_m] + \phi_m^{t2} \sin I \cos I  \bigg] \, .
\end{equation}
To a good approximation, this is equal to 

\begin{equation}
    \zeta_m \approx \frac{\bar{A} \Omega_o}{{\cal L}_m} Im[{\cal S}_{11}] \phi_m^{t2} \sin I \, , 
\end{equation}
and is equivalent, in our notation, to the expression given in Equation (70) of \citet{baland17}, where they have made the further approximation ${\cal L}_m\approx \bar{A} \Omega_o \phi_m$. 

Provided $Q>10$, $\zeta_m^{t,e}\gg\zeta_m^{t,a}$.  For a small or no inner core, $\bar{A}_s \ll \bar{A}_f$, ${\cal S}_{14} \ll {\cal S}_{12}$ and the phase lag can be approximated by 

\begin{equation}
    \zeta_m \approx \frac{\bar{A} \Omega_o}{{\cal L}_m} \bigg[ Im[{\cal S}_{11}] \phi_m^{t2} \sin I + Im[\tilde{m}_f] \left( \frac{\bar{A}_f}{\bar{A}} \frac{\Omega_p}{\Omega_o} \cos I - \frac{\phi_m}{e} Re[{\cal S}_{12} ]\right) \bigg]\, . \label{eq:predzetam}
\end{equation} 
The term proportional to $Im[\tilde{m}_f]$ captures the contribution to the phase lag from the out-of-plane component of the spin vector of the fluid core. It involves the same factor as in the prediction for the obliquity in Equation (\ref{eq:oblpredmf}). If the global elastic deformations caused by the misaligned fluid core are neglected, $Im[\tilde{m}_f]$ contributes to a positive phase lag.  But since $\frac{\phi_m}{e} Re[{\cal S}_{12}] > \frac{\bar{A}_f}{\bar{A}} \frac{\Omega_p}{\Omega_o} \cos I$, $Im[\tilde{m}_f]$ actually contributes to a negative phase lag (i.e. a phase lead).  For a large inner core, terms that involve $Im[\tilde{n}_s]$ are also important, and so are the global deformations captured by the compliance ${\cal S}_{14}$.  Just like for the prediction of the obliquity, a proper prediction of the phase lag must include global deformations induced by $\tilde{m}_f$ and $\tilde{n}_s$.

\section{Results}

\subsection{Viscous dissipation}

We first investigate the dissipation due to viscous coupling at the CMB and ICB in isolation.  EM coupling is turned off and the imaginary parts of all compliances are set to zero.  The real parts of compliances are retained so elastic deformations are part of the solutions, but there are no anelastic deformations and so no tidal dissipation. The parameterization of the viscous coupling constants $K_{cmb}$ and $K_{icb}$ is the same as that used in D21 (based on \cite{mathews05}), 

\begin{subequations}
 \begin{align}
 K_{cmb} &= \frac{\pi \rho_f r_f^4}{\bar{A}_f} \sqrt{\frac{\nu}{2\Omega_o}} \Big( 0.195 - 1.976 i \Big) \, ,
 \label{eq:Klamcmb} \\
 K_{icb} &= \frac{\pi \rho_f r_s^4}{\bar{A}_s} \sqrt{\frac{\nu}{2\Omega_o}} \Big( 0.195 - 1.976 i \Big) \, ,
 \label{eq:Klamicb}
 \end{align}
 \label{eq:Klam}
 \end{subequations}
 where $\nu$ is the kinematic viscosity.  These expressions are valid provided the flow in the boundary layer remains laminar.  As detailed in D21, the boundary layer flow is expected to be in a turbulent regime.  We take the same simple approach as that taken in D21; we use the above laminar model with the understanding that $\nu$ represents an effective turbulent viscosity.

\begin{figure}
\centering
    \includegraphics[width=15cm]{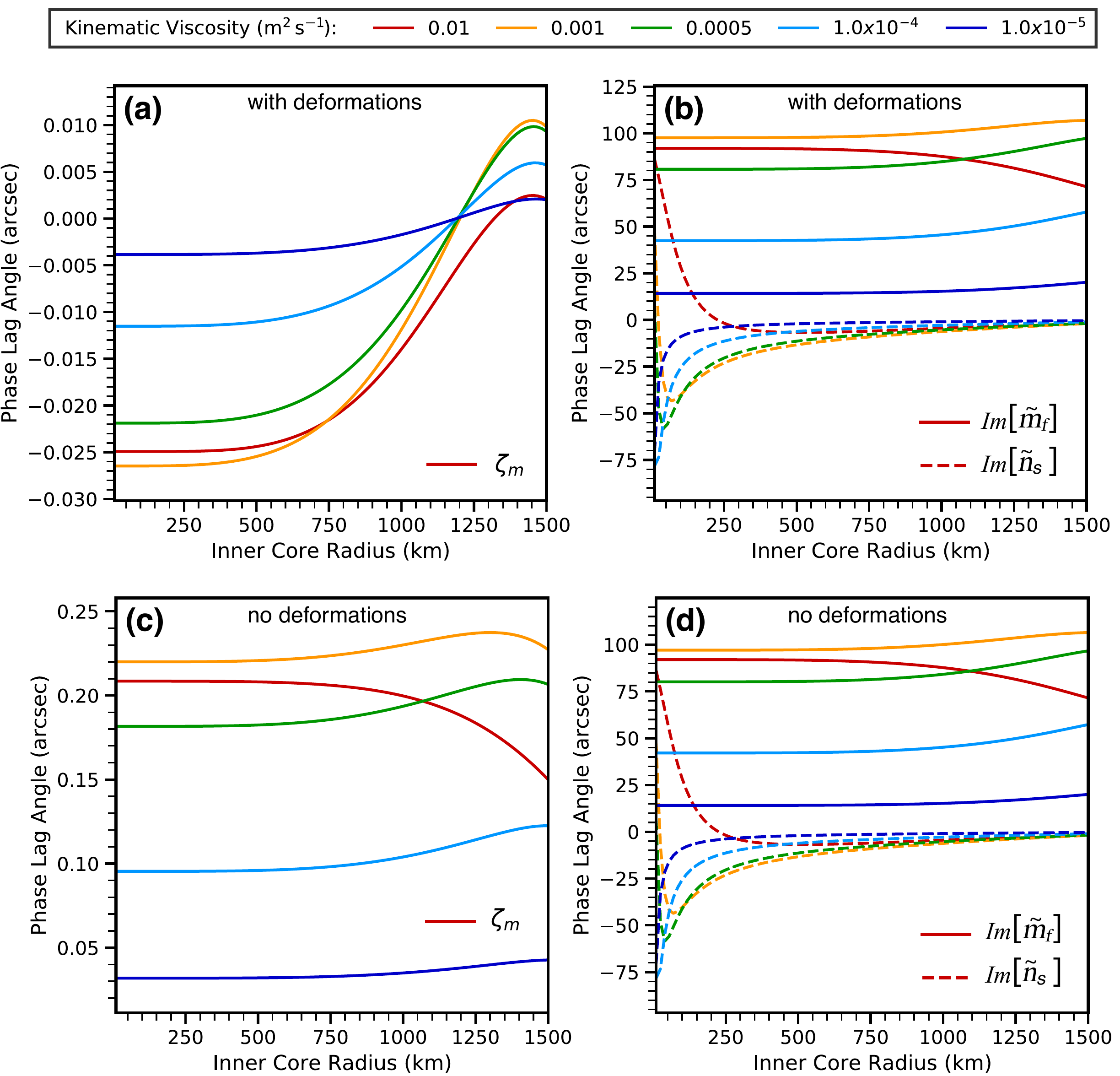}
    \caption{\label{fig:visc_obl} a) Phase lag of the mantle spin axis ($\zeta_m$)  b) fluid core spin axis ($Im[\tilde{m}_f]$, solid lines) and inner core symmetry axis ($Im[\tilde{n}_s]$, dashed lines) as a function of inner core radius and for different choices of kinematic viscosity (colour in legend). c) and d) idem, but with no deformations (all compliances ${\cal S}_{ij}$ set to zero).}
\end{figure}

Figure \ref{fig:visc_obl}ab shows how the mantle phase-lag $\zeta_m$ and the imaginary parts (out-of-plane components) of $\tilde{m}_f$ and $\tilde{n}_s$ vary as a function of inner core radius for different choices of the kinematic viscosity, $\nu$.  Let us first concentrate on results for a small inner core (radius $< 500$ km).  $\zeta_m$ is negative for all choices of $\nu$: the spin axis of the mantle is ahead of the Cassini plane (a phase lead).  The spin axis of the fluid core lags behind the Cassini plane ($Im[\tilde{m}_f]>0$). Starting from $\nu = 10^{-5}$ m$^2$ s$^{-1}$, viscous dissipation increases with increasing $\nu$, which leads to an increase in the magnitudes of $\zeta_m$ and $Im[\tilde{m}_f]$.  The dissipation peaks to a maximum value when $\nu$ is approximately equal to $10^{-3}$ m$^2$ s$^{-1}$.  With a further increase in $\nu$ beyond this value, viscous dissipation decreases, and so do the magnitudes of $\zeta_m$ and $Im[\tilde{m}_f]$.  

The peak in dissipation is connected to the viscous torque at the CMB,  proportional to $\sqrt{\nu} \, \tilde{m}_f$.  In the Cassini state equilibrium, with weak or no viscous coupling, the obliquity of the spin axis of the fluid outer core, $Re[\tilde{m}_f]$, is offset from the mantle by approximately 4 arcmin (see Figures 4 and 5 of D21). For a very small $\nu$, the viscous torque is weak, and so is the resulting viscous dissipation.  As $\nu$ is increased, $Re[\tilde{m}_f]$ is reduced; the spin axis of the fluid core is brought into an alignment with the mantle's rotation (see Figure 5 of D21).  When $\nu$ is very large, the differential velocity at the CMB is very small and, consequently, viscous dissipation is also weak.  The dissipation is then maximized when $\nu$ is sufficiently large to generate a large viscous torque, yet not so large as to prevent a misalignment between the spin axes of the fluid core and mantle.  For $\nu \approx 10^{-3}$ m$^2$ s$^{-1}$, which optimizes viscous dissipation, the mantle phase lead is $\sim 0.027$ arcsec and the fluid core phase lag is $\sim 100$ arcsec ($\sim 1.7$ arcmin).

Our results for a small or no inner core differ from those obtained by \citet{peale14}.  First, we use a different parameterization of the viscous torque, so for the same choice of $\nu$ the numerical values of the out-of-plane components of the mantle and fluid core spin orientations that we obtain are different.  But our results are also qualitatively different: in contrast to \citet{peale14}, we find that the net effect of viscous coupling at the CMB is to generate a mantle phase lead instead of a phase lag. The reason for this difference can be understood from the prediction of the mantle phase lag given by the approximate solution of Equation (\ref{eq:predzetam}) in section 2.5.  As we explained in that section, the solar torque acting on the elastic deformations induced by the out-of-plane component of the fluid core spin axis (through the compliance ${\cal S}_{12}$)  acts akin to a tidal torque.  This contribution to $\zeta_m$ is opposite to that caused by the viscous torque at the CMB and larger in magnitude.  As a result, the net effect of viscous coupling at the CMB is to generate a mantle phase lead.   Figure \ref{fig:visc_obl}cd shows how the results are altered when all compliances are set to zero (no deformations).  The mantle phase lag is now positive, consistent with the results of \citet{peale14}, and is increased in magnitude by approximately a factor 10.

As observed in Figure \ref{fig:visc_obl}ab, when the inner core radius exceeds 500 km, its presence alters the resulting mantle phase lead, reducing its magnitude. For a very large inner core, $\zeta_m$ can be positive (a mantle phase lag), with a magnitude peaking at 0.01 arcsec.  The influence of the inner core on $\zeta_m$ occurs through several mechanisms, as discussed in section 2.5.  First, as shown in Figure \ref{fig:visc_obl}b, the viscous torque at the ICB entrains a phase lead of the inner core spin axis (recall that the spin and symmetry axes of the inner core are virtually in alignment, $\tilde{n}_s \approx \tilde{m}_s$). This induces a gravitational torque on the mantle which contributes to a mantle phase lead (the contribution from the term $\zeta_m^{L,c}$ in the prediction given by  Equation \ref{eq:zetam1}).  The solar torque acting on the tilted inner core (the term $\zeta_m^{t,s}$) and the elastic deformation resulting from the latter (the term $\zeta_m^{t,e}$) both contribute to a phase lag.  These latter two contributions are more important than that from the gravitational torque, so the net effect of viscous coupling at the ICB is to generate a mantle phase lag.  When the inner core radius is $>500$ km, the magnitude of the net mantle phase lead (from viscous coupling at the CMB) is reduced.  For a very large inner core, the net effect from viscous coupling at both the CMB and ICB is a mantle phase lag.  

Just as elastic deformations induced by the out-of-plane component of the fluid core spin cannot be neglected, those induced by the out-of-plane component of the inner core tilt cannot either.  A convenient way to demonstrate this is to write the total perturbation in the moment of inertia produced by an inner core tilt in the form $\bar{A}_s \alpha_3 e_s (1 + k_s) \tilde{n}_s$, where $k_s$ is the equivalent of a Love number, capturing the added contribution to the change in moment of inertia induced by deformations (see Appendix C).  $k_s$ depends on inner core size and the rheology of the solid regions.  The sum of the contributions $\zeta_m^{t,s}$ and $\zeta_m^{t,e}$ from the inner core can then be written as $\zeta_m^{t,s}(1 + k_s)$.  For a rheology that is constrained to match $k_2=0.55$, $k_s$ falls between 0.6 and 0.9 (see Figure C.1). Hence, elastic deformations cannot be neglected in the prediction of $\zeta_m$. The contrast in the results of Figures \ref{fig:visc_obl}ab and \ref{fig:visc_obl}cd indeed illustrates the importance of including elastic deformations induced by the misaligned fluid core and inner core in the prediction of $\zeta_m$. (Note though that the solutions for $Im[\tilde{m}_f]$ and $Im[\tilde{n}_s]$ are virtually unchanged; these solutions are not altered significantly by elastic deformations.)

In summary, viscous coupling at the CMB and ICB generate a mantle phase lead for a small inner core, and a mantle phase lag for a large inner core.  As argued in D21, a conservative upper bound for the effective turbulent viscosity is $\nu \approx 5 \times 10^{-4}$ m$^2$ s$^{-1}$. This places an upper limit of 0.02 arcsec on the mantle phase lead. The out-of-plane components of the spin axes of the fluid and solid cores are substantially larger.   The spin axis of the fluid core lags behind the Cassini plane, with a maximum phase lag that can approach 100 arcsec. The inner core leads ahead of the Cassini plane, with a phase lead of a few 10s of arcsec for a small inner core, and limited to a few arcsec for a large inner core.  Note that these amplitudes are of the same order as their in-plane components (see Figure 5 of D21).

\subsection{Electromagnetic dissipation}

We now investigate dissipation caused by EM coupling.  We set viscous coupling to zero and again set the imaginary parts of all compliances to zero.  The differential velocity at the CMB and ICB shears the local radial magnetic field $B_r$.  This induces a secondary magnetic field which leads to a tangential force resisting the differential motion. This magnetic ``friction'' depends on the radial magnetic field strength $B_r$ and the electrical conductivity $\sigma$ on either side of the boundary \cite[][]{rochester60,rochester62,rochester68}.  

As argued in section 3.4 of D21, at the CMB of Mercury, EM coupling is expected to be much weaker than viscous coupling.  For simplicity, we simply assume no EM coupling at the CMB ($K_{cmb}=0$) and concentrate our efforts on the dissipation induced by EM coupling at the ICB.  We follow D21 and assume a parameterization for $K_{icb}$ given by 

 \begin{equation}
 K_{icb} = \frac{5}{4}(1-i){\cal F}_{icb} \left<B_r\right>^2 \, ,\label{eq:Kem}
 \end{equation}
 where $\left<B_r\right>$ is the r.m.s. strength of the radial component of the field at the ICB and
 
 \begin{equation}
{\cal F}_{icb} = \frac{\sigma \delta}{\Omega_o \rho_s r_s} \, , \label{eq:FM}
 \end{equation}
where $\sigma$ is the electrical conductivity (assumed equal in the fluid and solid core) and $\delta=\sqrt{2/(\sigma \mu \Omega_o)}$ is the magnetic skin depth, with $\mu=4\pi \times 10^{-7}$ N A$^{-1}$ the magnetic permeability of free space. We use $\sigma=10^6$ S m$^{-1}$, a reasonable value for Mercury's core \cite[e.g.][]{berrada21}. This parameterization is valid provided EM coupling remains in a weak-field regime which, as detailed in D21, is a reasonable assumption for Mercury.

\begin{figure}
\centering
    \includegraphics[width=15cm]{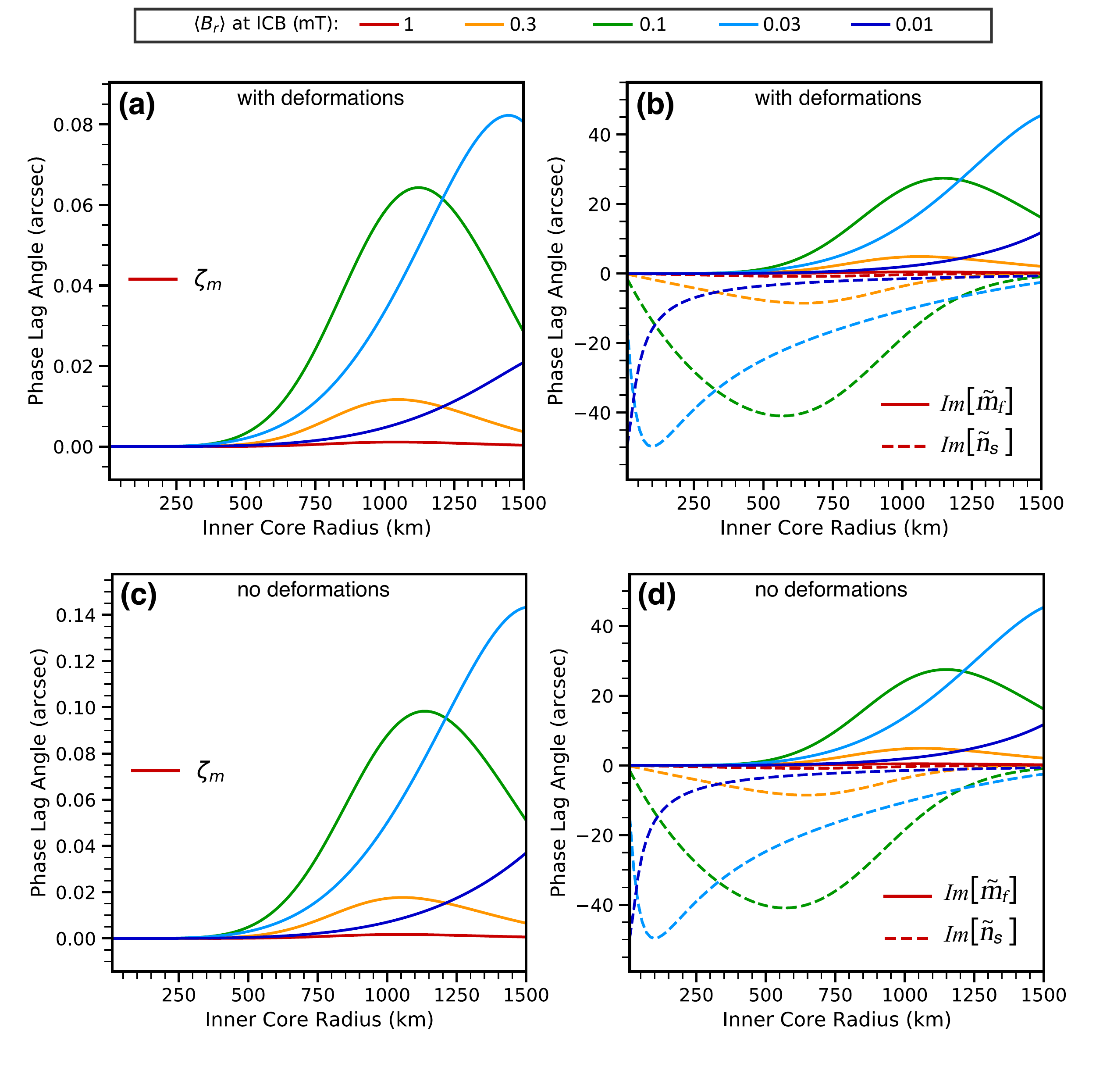} 
    \caption{\label{fig:br_obl} 
    a) Phase lag of the mantle spin axis ($\zeta_m$)  b) fluid core spin axis ($Im[\tilde{m}_f]$, solid lines) and inner core symmetry axis ($Im[\tilde{n}_s]$, dashed lines) as a function of inner core radius and for different choices of $\left<B_r\right>$ at the ICB (colour in legend). c) and d) idem, but with no deformations (all compliances ${\cal S}_{ij}$ set to zero).}
\end{figure}

Figure \ref{fig:br_obl}ab shows how $\zeta_m$ and the imaginary parts of $\tilde{m}_f$ and $\tilde{n}_s$ vary as a function of inner core radius for different choices of $\left<B_r\right>$.  The net effect of EM coupling at the ICB is to generate a mantle phase lag ($\zeta_m > 0$).  The EM torque (and dissipation) increases with the size of the inner core; the resulting mantle phase lag remains small ($<0.01$ arcsec) for an inner core radius $<500$ km.  For a large inner core, the magnitude of $\zeta_m$ can be considerably larger than that from viscous coupling, as high as $\sim0.08$ arcsec for $\left<B_r\right> =0.03$ mT.  

The EM torque is proportional to $\left<B_r\right>^2 (\tilde{m}_s - \tilde{m}_f)$.  EM dissipation is weak when $\left<B_r\right>$ is small, and also weak when $\left<B_r\right>$ is large, as then a strong EM coupling prevents a large differential rotation at the ICB (i.e. $\tilde{m}_s \approx \tilde{m}_f$).  Hence, just as for viscous coupling, EM dissipation is characterized by a saturation effect; it is maximized when $\left<B_r\right>$ is sufficiently large to generate a large EM torque but not too large as to prevent differential rotation.  This maximum dissipation is produced when $\left<B_r\right>$ is of the order $0.03-0.1$ mT and also depends on inner core size.

The spin axis of the fluid inner core lags behind the Cassini plane, while the spin axis of the inner core is displaced ahead of it.  The amplitude of their offsets is of the order of a few 10s of arcsec.  The inner core phase lead results in a mantle phase lag for the same reasons as explained in the previous section; the gravitational torque by the inner core generates a mantle phase lead, but the solar torque acting on the tilted inner core and the global deformations that it entrains produce a phase lag, and the latter contribution is larger in magnitude.  

As in the case of viscous coupling, elastic deformations induced by both the misaligned fluid core (through the compliance ${\cal S}_{12}$) and inner core (through ${\cal S}_{14}$) have a first order influence on the prediction of $\zeta_m$.  To illustrate this, Figure \ref{fig:br_obl}cd shows how the results are altered when all compliances are set to zero.  The solutions for $\zeta_m$ are qualitatively similar, but their amplitudes are different.  Note again that, as observed in the case of viscous coupling, the solutions for $Im[\tilde{m}_f]$ and $Im[\tilde{n}_s]$ are not altered significantly by elastic deformations.

In summary, EM coupling at the ICB generates a mantle phase lag which, for an inner core radius of 1000 km or larger, can be as high as 0.08 arcsec for a $B_r$ field close to 0.03 mT that optimizes dissipation.  Such a field strength is a factor 100 larger than the field measured at Mercury's surface \cite[e.g.][]{anderson12}, but it is not an unreasonable estimate if the field geometry deep within the core is dominated by small length scales \cite[e.g.][]{christensen06b}.  Hence, it may well be that dissipation at the ICB from EM coupling is close to its optimal value at present-day.  If the inner core radius is 1000 km or larger, the mantle phase lag resulting from EM coupling at the ICB is substantially larger than the maximum phase lag or lead generated by viscous coupling.

\subsection{Tidal dissipation}

We now turn to the dissipation resulting from anelastic deformations.  To isolate their effect on the mantle phase lag, we set both viscous and EM coupling to zero.  The delayed, anelastic response of Mercury to tidal forces depends on the ratio $k_2/Q$ which, in our formulation, is captured by the imaginary component of the compliance ${\cal S}_{11}$ (see Equation \ref{eq:k2Q}).  We do not prescribe values of $Q$; instead, we specify  the viscosity of each solid regions, and calculate the resulting $Q$ on the basis of $Im[{\cal S}_{11}]$.  We recall that we assume a Maxwell rheology in solid regions, see Appendix C for the computation of the compliances.  Global anelastic deformations also occur in response to the pressure force at the CMB from the misaligned fluid core spin axis and from the gravitational force induced by a tilted inner core.  These are captured by the imaginary parts of the compliances ${\cal S}_{12}$ and ${\cal S}_{14}$, respectively.

\begin{figure}[h]
\centering
    \includegraphics[height=8cm]{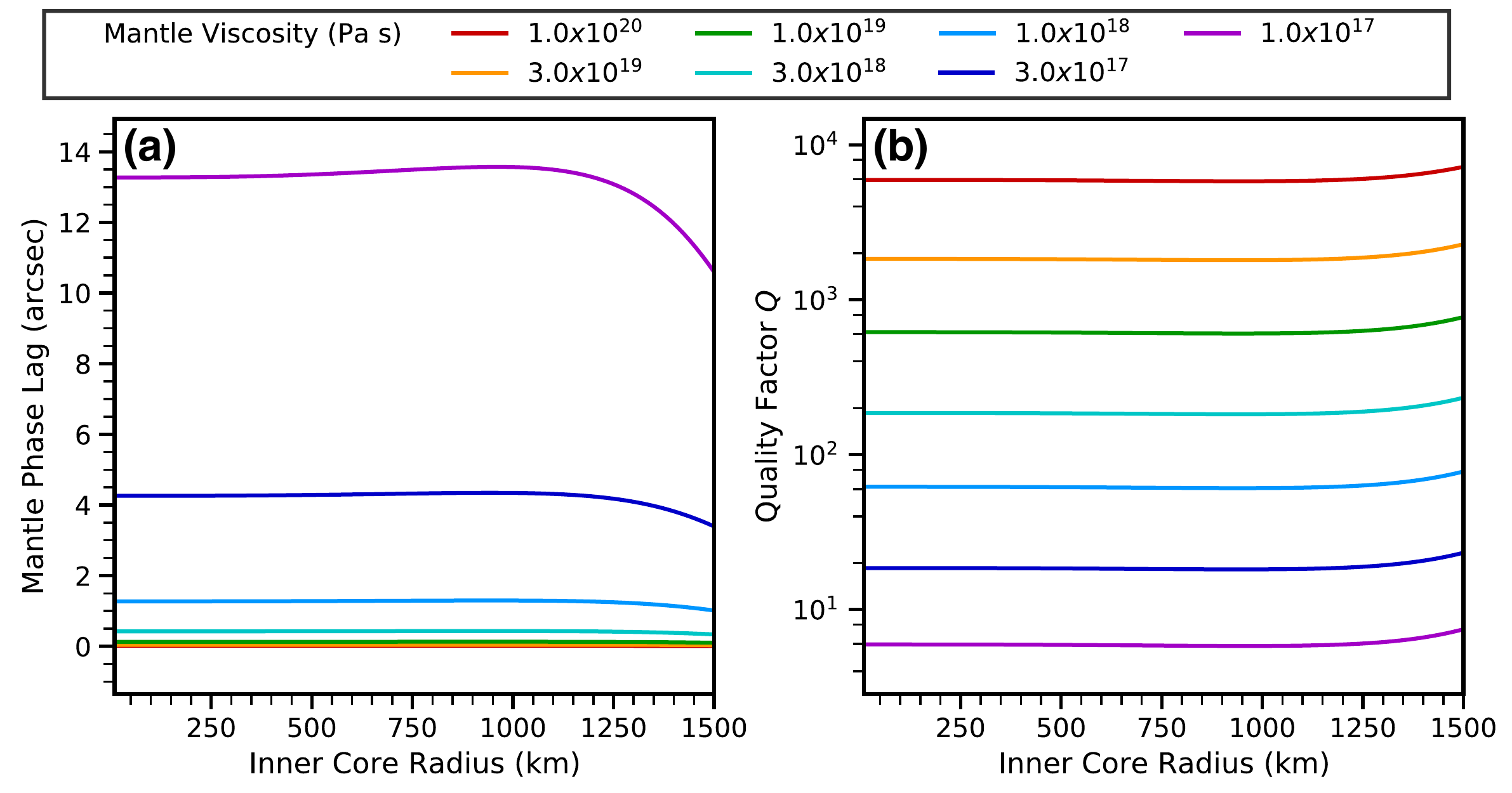} 
    \caption{\label{fig:manvisc} (a) Phase lag of the mantle $\zeta_m$ and (b) tidal quality factor $Q$ as a function of inner core radius and for different choices of mantle viscosity.}
\end{figure}

Figure \ref{fig:manvisc} shows how $\zeta_m$ and $Q$ vary as a function of inner core radius for different choices of mantle viscosity; these values refer to the bulk viscosity of the whole of the mantle.  In all cases, the inner core viscosity is fixed at $10^{20}$ Pa s. Tidal deformations result in a positive $\zeta_m$, in other words a mantle phase lag. For our largest choice of mantle viscosity, $10^{20}$ Pa s, $Q$ is approximately 6000 and the phase lag is very small, approximately 0.01 arcsec.  As the viscosity of the mantle is decreased, $Q$ is reduced and the phase lag increases in amplitude.  An approximate empirical relationship between $\zeta_m$ and $Q$ based on our results is $\zeta_m \sim (80/Q)$ arcsec.  When $Q$ is of the order of 100, the mantle phase-lag is of the order of 1 arcsec, consistent with the results obtained by \citet{baland17}. Unless $Q$ is larger than a few hundred, the deviation of the mantle spin axis from the Cassini plane caused by anelastic deformations is significantly larger in magnitude than that from EM and viscous coupling at the fluid core boundaries.

\begin{figure}[h]
\centering
    \includegraphics[height=8cm]{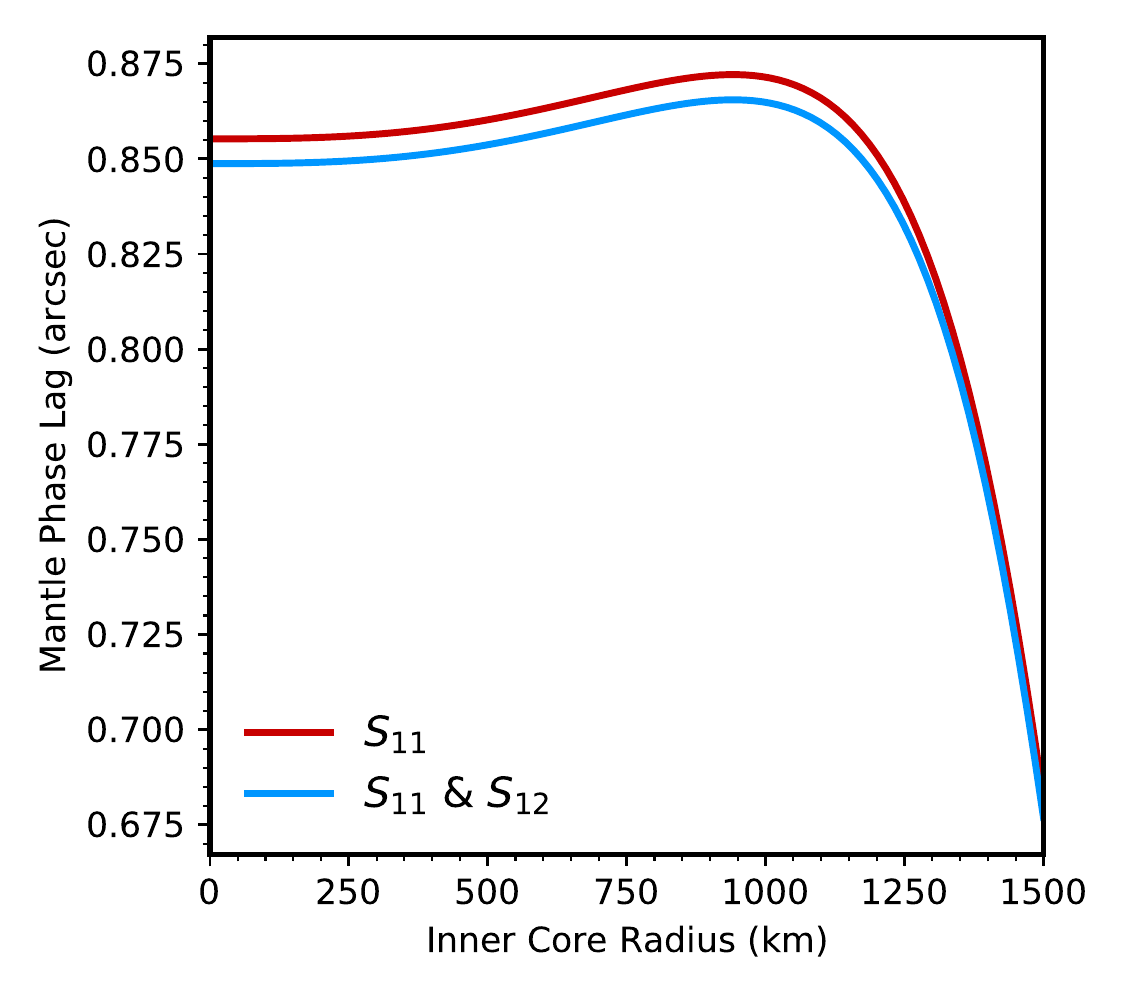} 
    \caption{\label{fig:compliances} Phase lag of the mantle $\zeta_m$ as a function of inner core radius when only the imaginary part(s) of ${\cal S}_{11}$ (solid red line) and, ${\cal S}_{11}$ and ${\cal S}_{12}$ (solid blue line) are retained.}
\end{figure}

For all cases in Figure \ref{fig:manvisc}a, the dominant contribution to $\zeta_m$ is from tidal dissipation (the term $\zeta_m^t$ in Equation \ref{eq:zetam1}).  There is a small secondary contribution (of the order of 1\%) from the term $\zeta_m^{t,a}$, the delayed anelastic response of the mantle to the misaligned obliquity of the fluid core and inner core.  In the absence of viscous and EM coupling, $Re[\tilde{m}_f]$ is of the order of 4 arcmin, while $Re[\tilde{n}_s]$ is very small, approximately 1.5 arcsec (see Figure 4 of D21), so it is predominantly the part from $Im[{\cal S}_{12}] Re[\tilde{m}_f]$ that contributes to $\zeta_m^{t,a}$.  To illustrate this, Figure \ref{fig:compliances} shows an example of how the solution for $\zeta_m$ as a function inner core radius differs when only ${\cal S}_{11}$ is retained, versus when both ${\cal S}_{11}$ and ${\cal S}_{12}$ are retained, with all other compliances set to zero.  For these solutions, the bulk viscosity of the mantle is set to $1.5\times 10^{18}$ Pa s and gives a $Q$ of approximately 100. The delayed, anelastic response of the mantle to the pressure force at the CMB does affect the resulting mantle phase lag, but it only reduces it by a small amount (not more than 0.0125 arcsec on Figure \ref{fig:compliances}).  When all other compliances are included, the solutions is virtually identical to that shown in Figure \ref{fig:compliances} when only ${\cal S}_{11}$ and ${\cal S}_{12}$ are retained.

\begin{figure}[h]
\centering
    \includegraphics[height=8cm]{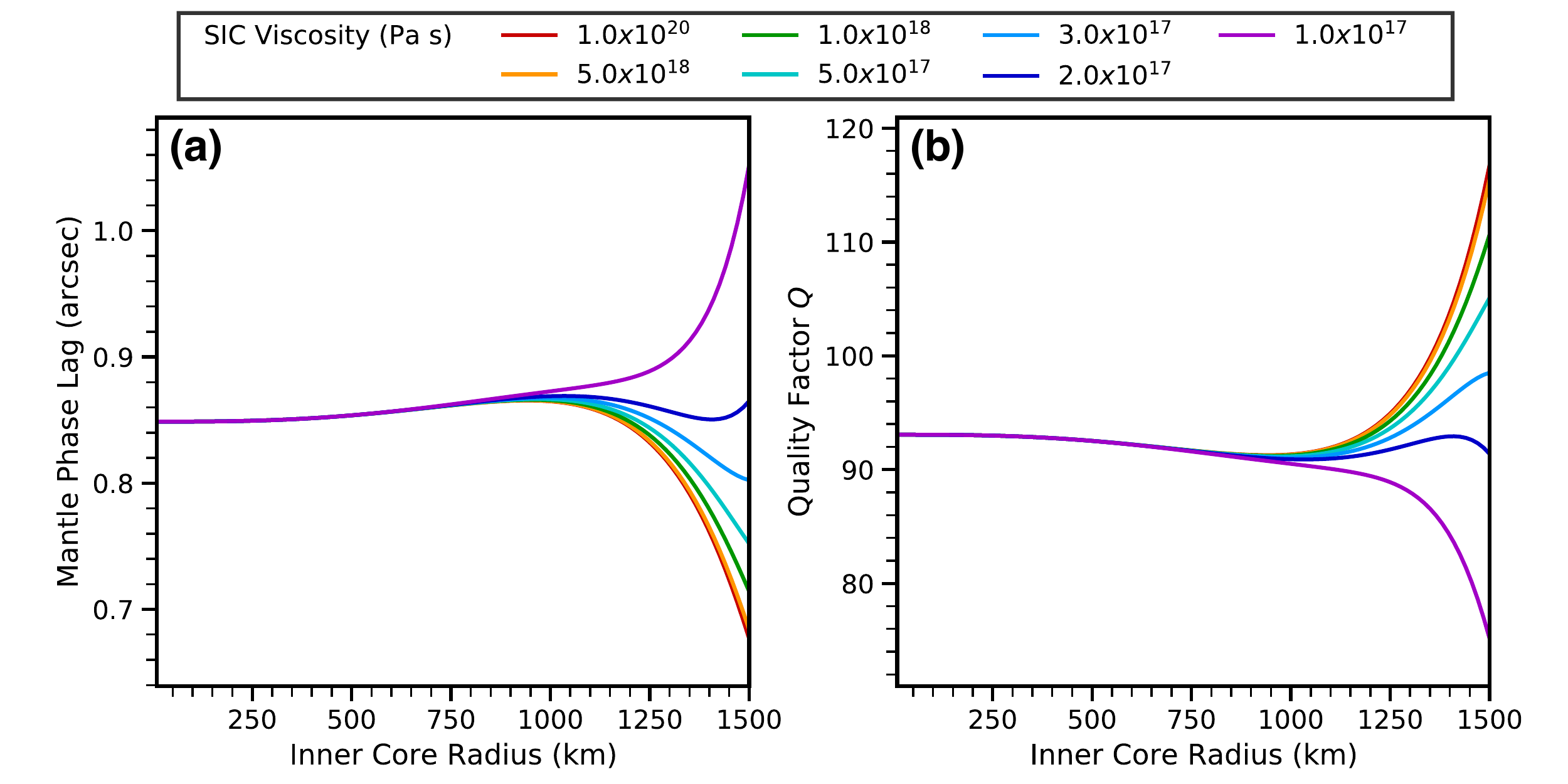} 
    \caption{\label{fig:sicvisc} (a) Phase lag of the mantle $\zeta_m$ and (b) tidal quality factor $Q$ as a function of inner core radius and for different choices of solid inner core viscosity.}
\end{figure}

Both $Q$ and $\zeta_m$ are affected by the size of the inner core.  This can be observed in Figure \ref{fig:manvisc} but is better highlighted by Figure \ref{fig:compliances}. A large, stiff inner core implies smaller global anelastic deformations, leading to an increase in $Q$ with inner core size, and a decrease in $\zeta_m$. In turn, the inner core viscosity can also influence $Q$ and the resulting $\zeta_m$.  Keeping the mantle viscosity fixed at $1.5\times 10^{18}$ Pa s, Figure \ref{fig:sicvisc} shows how $\zeta_m$ and $Q$ vary as a function of inner core radius for different choices of inner core viscosity.  Provided the inner core radius is smaller than approximately 1000 km, the inner core viscosity has a negligible influence on $Q$ and $\zeta_m$.  However, for a large inner core (radius $>1000$ km) and a low viscosity ($<10^{17}$ Pa s), $Q$ can be substantially reduced and $\zeta_m$ substantially increased.  Note that the empirical relation $\zeta_m \sim (80/Q)$ arcsec remains applicable in all cases shown in Figures \ref{fig:manvisc}-\ref{fig:sicvisc}. 

\section{Discussion}

We have shown that viscous coupling at the CMB results in a mantle phase lead ahead of the Cassini plane, while viscous and/or EM coupling at the ICB results in a mantle phase lag.  Elastic deformations induced by the misaligned spin axes of the fluid core and inner core play a first order role in the resulting mantle phase $\zeta_m$.  The influence on $\zeta_m$ from EM and viscous coupling at the ICB gets proportionally more important the larger the inner core is.  The net phase that results from dissipation at both the CMB and ICB depends then on the inner core size and on the parameters on which the viscous and EM torques depend, notably, on the kinematic viscosity and the amplitude of the radial magnetic field at the ICB. Importantly, a saturation effect limits the dissipation  and thus the maximum phase lead or lag that can be generated by either viscous and EM drag.

Overall, viscous and EM coupling at the fluid core boundaries generate only a small deviation of no more than 0.1 arcsec of the mantle spin away from the Cassini plane.  This is a factor 10 smaller than the smallest measurement error on the different estimates of the mantle spin position, which is $\sim1$ arcsec (see Table \ref{tab:lag}).  Hence, unless measurement errors can be reduced by more than a factor 10, it is unlikely that observations of the mantle phase lag can yield useful constraints on inner core size and/or viscous and EM coupling at the fluid core boundaries. 

From our results shown in Figures \ref{fig:visc_obl} and \ref{fig:br_obl}, we can compute the dissipation at the CMB and ICB, respectively, from

\begin{subequations}
\begin{equation}
Q_{cmb} = \Omega_o^3 \bar{A}_f Im[K_{cmb}] \, \big|  \tilde{m}_f \big|^2 \, , 
\end{equation}
\begin{equation}
Q_{icb} = \Omega_o^3 \bar{A}_s Im[K_{icb}] \, \big|  \tilde{m}_f - \tilde{m}_s \big|^2 \, .
\end{equation}
\end{subequations}
At its peak value, the dissipation from viscous coupling at the CMB is approximately $2.5 \times 10^7$ W, and that at the ICB is $5 \times 10^6$ W.  Expressed in terms of heat fluxes, these correspond to $q_{cmb}= Q_{cmb}/4\pi r_f^2 \approx 5 \times 10^{-7}$ W m$^{-2}$ and $q_{icb}= Q_{icb}/4\pi r_s^2 \approx 2 \times 10^{-7}$ W m$^{-2}$ (the latter based on an inner core radius of $r_s=1500$ km). The peak dissipation at the ICB from EM coupling is approximately $6 \times 10^6$ W ($q_{icb}\approx 2 \times 10^{-7}$ W m$^{-2}$).  These are small compared to estimates of the heat flow out of the core, which are of the order of $10^{11}$ W, corresponding to a heat flux of $2 \times 10^{-3}$ W m$^{-2}$ \cite[e.g.][]{knibbe18,tosi13,grott11}.   Dissipation at the CMB and ICB from viscous and/or EM coupling contributes to only a very small fraction of the internal heat budget of Mercury.  Furthermore, the heat released at the ICB from viscous and EM dissipation is very small compared to the latent heat associated with inner core growth \cite[of the order of $10^{11}$ W, e.g.][]{knibbe18} and adds a negligible contribution to the convective power in Mercury's fluid core and to the power required to generate its dynamo. 

Tidal dissipation generates a mantle phase lag with a magnitude inversely proportional to the quality factor $Q$. An approximate empirical relationship derived from our results is $\zeta_m \sim (80/Q)$ arcsec.  For $Q$ of the order of $100$, the phase lag is approximately 1 arcsec. Unless $Q>1000$, the phase lag produced by tidal dissipation dominates that due to viscous and EM coupling at the fluid core boundaries. $Q$ is proportional to the bulk mantle viscosity; a $Q$ value of 100 corresponds to a bulk mantle viscosity of approximately $10^{18}$ Pa s, based on a Maxwell rheology. 

Thermal evolution and mantle convection models tuned to match Mercury's history of magmatism and radial contraction tend to favor a stiff mantle with high viscosities in the range of $10^{19}-10^{22}$ Pa s \cite[][]{grott11,tosi13,michel13,ogawa16,knibbe18}.  A high mantle viscosity is also required to maintain deep
seated mass anomalies so as to explain Mercury's long wavelength topography \cite[][]{james15} and nonhydrostatic shape \cite[][]{matsuyama09}.  Based on these, we expect then a small phase lag angle of the order of 0.1 arcsec or smaller from tidal dissipation.  However, the viscosity of the lower mantle that is compatible with observations of $k_2$ falls in the range of $10^{13}-10^{18}$ Pa s \cite[e.g.][]{steinbrugge21}.  The viscosity in the top part of the mantle is expected to be higher, as temperature decreases with radius, so a bulk viscosity of $10^{18}$ Pa s in order to fit $k_2$ may not be unreasonable.  If so, the phase lag from tidal dissipation can be expected to be of the order of 1 arcsec.  

The precession of the pericentre causes a deviation of the spin pole from the Cassini plane equivalent to a phase lag of 0.85 arcsec \cite[][]{baland17}.  With a $Q$ of approximately 80, we expect the net phase lag of the spin pole to be $\sim1.85$ arcsec.  All measurements of the spin pole position listed in Table \ref{tab:lag} are consistent with this.  Even the measurement by \citet{mazarico14}, which suggests a phase lead of approximately 7.8 arcsec, remains within its error bar consistent with a small phase lag.  The largest possible phase lag allowed by the different spin pole measurements is approximately 12 arcsec.  This provides a lower bound for $Q$ in the vicinity of 10.  If we take the most recent measurement of \citet{bertone21} as a benchmark, the largest phase lag allowed by the measurement error is approximately 1.8 arcsec.  Removing the contribution from the precession of the pericentre, this leaves a maximum of 1 arcsec caused by tidal dissipation, elevating the lower bound for $Q$ to $\sim80$.

As these simple calculations show, an improved measurement of the mantle spin position can yield a constraint on $Q$, and in turn, on the mantle viscosity.  Lower bounds on $Q$ of 10 and 100 corresponds to lower bounds on the bulk mantle viscosity of $10^{17}$ and $10^{18}$ Pa s, respectively.  It is worth emphasizing that these viscosity values are based on a Maxwell rheology in the mantle.  Using an Andrade-pseudoperiod model, believed to capture better the rheology of planetary mantles \cite[e.g.][]{padovan14,steinbrugge21}, the viscosity would be higher for the same $Q$, so the values quoted above remain lower bounds.  As we have shown, a large inner core (radius $>1000$ km) with a bulk viscosity lower than $10^{17}$ Pa s can reduce the global $Q$ and increase the phase lag. A large inner core with a very low viscosity would then permit to achieve the same $Q$ with a higher bulk mantle viscosity, though the values quoted above remain lower bounds.

We have shown that the delayed, anelastic deformations caused by the pressure force at the CMB from the misaligned rotation vector of the fluid core contribute to the total mantle phase lag. However, this is a small contribution, of the order of 1\% compared to the anelastic response of the mantle to tidal forcing.  We note though that our results are based on a uniform mantle viscosity; the amplitude of this contribution may be increased if the viscosity is weakest at the bottom of the mantle -- which is indeed what we expect.  An improvement on our model would be to consider radial variations in the material properties in the mantle, in particular its viscosity.

\section{Conclusion}

In this study, we computed predictions of the deviation of Mercury's spin axis from the Cassini plane (out-of-plane component) from different dissipation mechanisms. Viscous coupling at the CMB results in a phase lead, viscous and EM coupling at the ICB produce a phase lag, and tidal dissipation produces a phase lag.  

The magnitude of the mantle phase lead or lag from viscous and EM coupling depends on the inner core size, the kinematic viscosity, and magnetic field strength, though it cannot exceed a maximum value.  For a small inner core, viscous drag at the CMB dominates and produces a maximum phase lead of 0.027 arcsec.  For a large inner core (radius $>$ 1000 km), EM drag at the ICB can exceed viscous coupling at both the ICB and CMB, and produces a phase lag that does not exceed 0.1 arcsec. For both viscous and EM coupling, the solar torque acting on the global elastic deformations induced by the out-of-plane components of the spin axes of the fluid core and inner core play a first order role in the resulting mantle phase.  Tidal dissipation in the mantle produces a phase lag with a magnitude inversely proportional to the quality factor $Q$. For a $Q$ of the order of 100, the phase lag is approximately 1 arcsec. 

Our results suggest that dissipation should not displace Mercury's mantle spin axis away from the Cassini plane by more than a few arcsec.  This is indeed in agreement with observations.  In turn, the limited phase lag suggested by observations ($\sim$1 to 10 arcsec) implies lower limits on $Q$ and the bulk mantle viscosity which cannot be much smaller than $10$ and $10^{17}$ Pa s, respectively.  A more precise measurement of the position of the spin axis can in principle provide a constraint on $Q$ and thus on the bulk mantle viscosity.

\appendix

\section{Calculation of the phase lag angle}

The classical Cassini State of Mercury is characterized by the co-planar precession of the orbit and spin poles of the planet about the Laplace pole.  The Cassini plane is  defined as the plane spanned by the axes of the orbit and Laplace poles (the normals to the orbital and Laplace planes, respectively).  If Mercury's spin pole were to obey a classical Cassini state exactly, it should lie in the Cassini plane.  Dissipation induces a misalignment of the spin pole away from the Cassini plane characterized by an angle of offset $\zeta_m$, defined positive and corresponding to a phase lag if it trails behind the Cassini plane.  Conversely, a negative $\zeta_m$ corresponds to a spin pole that is ahead of the Cassini plane and to a phase lead.  In this Appendix, we explain how we calculate the phase lag angles $\zeta_m$ and their errors that are listed in Table \ref{tab:lag} based on measurements of the orientation of the spin pole.

The orientation of the spin pole is given in terms of its right ascension ($\alpha$) and declination ($\delta$) angles with respect to the International Celestial Reference Frame (ICRF). The Cartesian components of a unit vector ${\bf u} = (u_x,u_y,u_z)$ pointing to a coordinate  ($\alpha,\delta$) on this imaginary celestial sphere are

\begin{equation}
    u_x = \cos(\delta) \cos(\alpha) \, , \quad u_y = \cos(\delta) \sin(\alpha)\, , \quad u_z = \sin(\delta) \, ,
\label{eq:RADECtoVec}
\end{equation}
where the $z$-axis is aligned with the celestial pole ($\delta = \frac{\pi}{2}$) and the $x$-axis is aligned with zero right ascension ($\alpha=0$).  The orientations of the Laplace pole ($\alpha_L,\,\delta_L$) and orbit pole ($\alpha_O,\,\delta_O$) at epoch J2000 are calculated in \cite{baland17}, and are

\begin{subequations}
\begin{align}
\alpha_L = (273.811048\pm0.324494)^\circ \, , \quad & \delta_L = (69.457475\pm0.259017)^\circ \, ,\\     
\alpha_O =  (280.987906\pm0.000009)^\circ \,  , \quad & \delta_O = (61.447794\pm0.000006)^\circ \, .    \label{eq:poles}
\end{align}
\end{subequations}

\begin{figure}[h]
\begin{center}
   \hspace*{-2cm} \includegraphics[width=20cm]{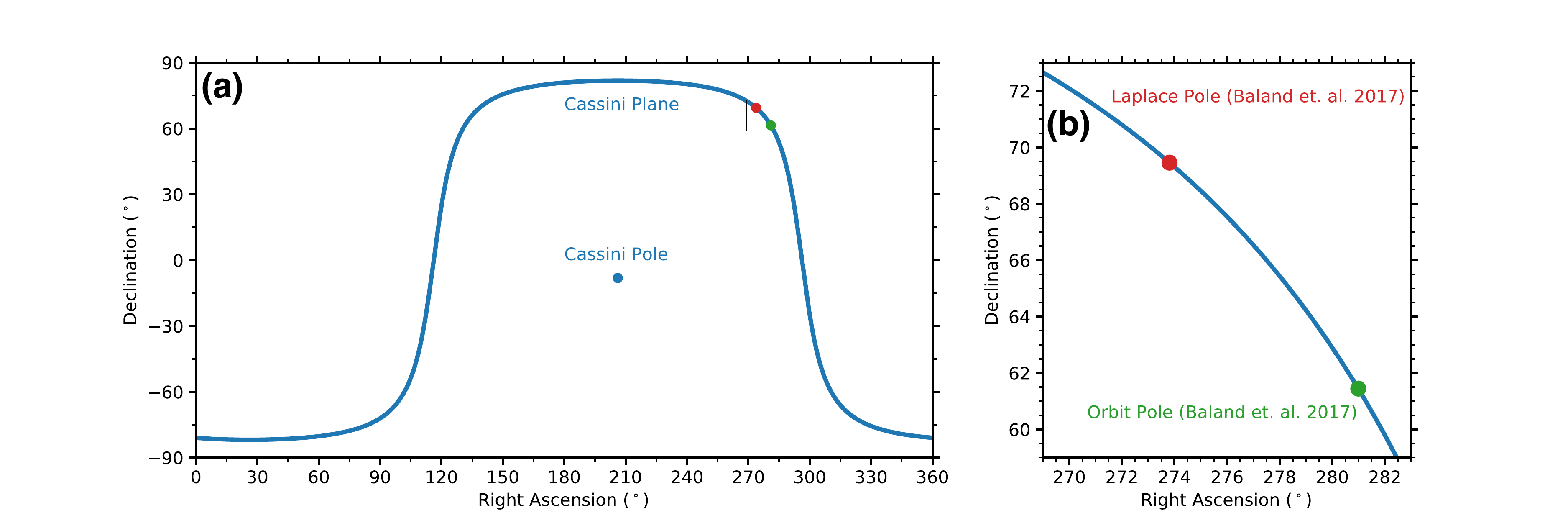} 
    \caption{\label{fig:cassinicircle} (a) The great circle of the Cassini plane on the celestial sphere as a function of right ascension and declination angles. (b) Close-up view in the vicinity of the Laplace and orbit poles. } 
\end{center}
\end{figure}

The unit vectors derived from the central values of the right-ascension and declination measurements of the Laplace and orbit poles are denoted with ${\bf u}_L$ and ${\bf u}_O$ respectively (these are denoted by $\mitbf{\hat{e}_3^L}$ and $\mitbf{\hat{e}_3^I}$, respectively, in the main text).  The Cassini plane corresponds to the plane that passes through the origin of the ICRF and whose great circle on the celestial sphere joins both the Laplace and orbit poles.  To define this great circle as a function of $\delta$ and $\alpha$, one must first determine the unit normal to the Cassini plane, defined by

\begin{equation}
{\bf u}_C = \frac{{\bf u}_L \times {\bf u}_O}{ \sqrt{1 - ({\bf u}_L\cdot{\bf u}_O)^2}} \, .
\end{equation}
The function of $\delta$ and $\alpha$ that defines the great circle can be found from the criteria that ${\bf u}_C \cdot {\bf u}=0$ with ${\bf u}$ defined as in Equation (\ref{eq:RADECtoVec}). In this manner, one can construct the great circle of the Cassini plane on the celestial sphere. Figure \ref{fig:cassinicircle}a shows how this great circle maps on a two dimensional projection of the celestial sphere. Figure \ref{fig:cassinicircle}b shows a close up view in the vicinity of the Laplace and orbit poles.

For a measurement of the spin pole orientation given as a pair ($\alpha$, $\delta$), its corresponding unit vector is denoted by ${\bf u}_S$. The phase lag angle, $\zeta_m$, between the great circle of the Cassini plane and the orientation of the spin pole is obtained from \cite[e.g. Eq. 41 of][]{baland17}

\begin{equation}
\sin(\zeta_m)  = \frac{{\bf u}_S \cdot({\bf u}_L \times {\bf u}_O)}{ \sqrt{1 - ({\bf u}_L\cdot{\bf u}_O)^2}} \, . \label{eq:lag}
\end{equation}
The numerical values for $\zeta_m$ given in Table \ref{tab:lag} in the main text are calculated from Equation (\ref{eq:lag}), using the central values of the the Laplace and orbit poles given in Equation (\ref{eq:poles}) and the central values of the spin pole measurements.  

The error in the phase lag is constructed from the errors in right ascension ($\Delta \alpha$) and declination ($\Delta \delta$).  For each spin pole measurement, an ellipse of error can be drawn around the central value. The phase lag error corresponds to the distance $\Delta \zeta_m$ between the central value and a point on this ellipse, in the direction perpendicular to the great circle of the Cassini plane.  We express this direction by an angle $\theta_o$ between ${\bf u}_C$ and the local unit vector in the direction of the increasing right ascension (${\bf \hat{\alpha}} = \-\hat{x} \sin \alpha_o + \hat{y} \cos \alpha_o$), at the location of the spin pole.  Graphically, on Figure \ref{fig:cassinicircle}b, $\theta_o$ corresponds to the angle between the x-axis and the direction perpendicular to the great circle of the Cassini plane at the location of the orbit pole.  We take $\alpha_o=281.0075^\circ$ as our reference spin pole position, which gives $\theta_o=17.17^\circ$. The distance $\Delta \zeta_m$ is then found by 

\begin{subequations}
\begin{equation}
\Delta \zeta_m = \sqrt{(\Delta x)^2 + (\Delta y)^2}
\end{equation}
where 

\begin{align}
    \Delta x & = 3600 \cdot \Delta \alpha \cdot \cos \theta_o \cdot \cos \delta_o \, ,\label{eq:dx}\\
    \Delta y & = 3600 \cdot \Delta \delta \cdot \sin \theta_o \, .
    \end{align}
\end{subequations} 
The factor 3600 converts degrees to arcseconds and the factor $\cos \delta_o$ in the expression for $\Delta x$ scales the angular error $\Delta \alpha$ at declination $\delta_o$ to its proper angular arc distance in right ascension.  We take $\delta_o=61.415^\circ$.

The phase lag errors calculated by this method are based solely on the uncertainty in the position of the spin pole at epoch J2000 reported in different studies.  Uncertainties in the determination of the Laplace pole, orbit pole and precession rate translate to an error in the precise location of the great circle of the Cassini plane on the Celestial sphere, both today and back at epoch J2000, and consequently to an additional error on the phase lag angle.   Depending on the method used to retrieve these orbital elements, at the location of the spin pole, this corresponds to a phase lag error of the order of 0.02 arcsec \cite[][]{baland17} to 0.2 arcsec \cite[][]{stark15b}. Spin pole measurements reported in different studies are made at different epochs (or more precisely over a time span with respect to a mean epoch) and not all studies give the details of how the projection back to epoch J2000 is carried out.  The phase lag error connected to the uncertainties in orbital elements may then be larger than 0.2 arcsec in individual studies.  Nevertheless, this error is typically an order of magnitude smaller than that connected to the spin pole positions reported in Table \ref{tab:lag} and we simply neglect it here.

For the same spin pole positions, the phase lags that we calculate in Table \ref{tab:lag} are slightly different than those given in Table C.2 of \citet{baland17}.  This is because of the choice made in the specific values of the Laplace pole.  Note also that our phase lag errors are smaller than those given in \citet{baland17}.  The method to calculate $\Delta \zeta_m$ is not detailed in \citet{baland17}, so the reason for this difference is unknown.  We note however that if the factor $\cos \delta_o$ is omitted in Equation (\ref{eq:dx}) the $\Delta \zeta_m$ that we obtain are closer to those given in Table C.2 of \citet{baland17}, so a part of the discrepancy may be due to this. We also note that estimates of $\alpha$ and $\delta$ are correlated in some studies, which causes the ellipse of error to be tilted in 2D plots like the one we show in Figure 2 of the main text. We do not take this tilt into account in our calculations of $\Delta \zeta_m$. Instead, we simply assume an ellipse with semi-major (semi-minor) axis equal to the largest (smallest) value between $\Delta \alpha \cos \delta_o$ and $\Delta \delta$.  

\section{Modification of the rotational model}

Tidal deformations of Mercury's figure occur in response to the imposed solar gravitational potential.  The deformations are of spherical harmonic degree 2 and hence induce a perturbation in the moment of inertia tensor.  The reshaping of Mercury's figure alters the amplitude of the solar torque acting on it and it also alters Mercury's angular momentum response.  For a purely elastic deformation, the tidal bulge is aligned with the line connecting the centre of Mercury to the Sun. Anelastic deformations from internal dissipation results in delayed response and to a misalignment of the tidal bulge.  The solar torque acting on the delayed part of the deformation is referred to as the tidal torque.  We show in this Appendix how the rotational model of D21 is modified to take into account viscoelastic deformations.   For brevity, we do not repeat the whole presentation of the model but only point out its modifications. All variable names and symbol that are not explicitly defined here are identical to those used in D21.

\subsection{Perturbation in the moment of inertia tensor}

As seen in the mantle frame, the inner core figure axis and the rotation vectors of the mantle, fluid core and inner core all precess in the retrograde direction.  The periodic changes in the gravitational and centrifugal potential associated with these lead to global deformations, and thus to a perturbation in the moment of inertia tensor of Mercury $\Delta \boldsymbol{{\cal I}}$.  These involve the off-diagonal terms $(\Delta \boldsymbol{{\cal I}})_{13}$ and $(\Delta \boldsymbol{{\cal I}})_{23}$.  In the complex notation used in D21, we write 
 
\begin{equation}
(\Delta \boldsymbol{{\cal I}})_{13}(t) + i (\Delta \boldsymbol{{\cal I}})_{23}(t) = \tilde{c} \, \exp[{i {\omega} \Omega_o t}] \, ,\label{eq:c123t}
\end{equation} 
where $\tilde{c}\equiv \tilde{c}({\omega} \Omega_o)$ is the amplitude of the perturbation at frequency ${\omega} \Omega_o$, where ${\omega}$ is given by Equation (\ref{eq:omega1}).  Equivalent definitions are used for the perturbation in the moment of inertia tensors of the fluid core and inner core, with $\tilde{c}_f$ and $\tilde{c}_s$ denoting their amplitudes, respectively.  The amplitudes $\tilde{c}$, $\tilde{c}_f$ and $\tilde{c}_s$ are expressed as a linear combination of the rotation variables and a set of compliances.  Following the notation introduced by \cite{buffett93}, we denote these compliances by ${\cal S}_{ij}$.   The perturbation in the moment of inertia tensors from internal contributions are defined as

\begin{subequations}
\begin{align}
\tilde{c}^{\,i} & = \bar{A} \big( {\cal S}_{11} \tilde{m}  + {\cal S}_{12} \tilde{m}_f + {\cal S}_{13} \tilde{m}_s + {\cal S}_{14} \tilde{n}_s \big) \, , \label{eq:ci1}\\
\tilde{c}_f^{\,i} & = \bar{A}_f \big( {\cal S}_{21} \tilde{m} + {\cal S}_{22} \tilde{m}_f + {\cal S}_{23} \tilde{m}_s + {\cal S}_{24} \tilde{n}_s \big) \, , \\
\tilde{c}_s^{\,i} & = \bar{A}_s \big( {\cal S}_{31} \tilde{m}  + {\cal S}_{32} \tilde{m}_f + {\cal S}_{33} \tilde{m}_s + {\cal S}_{34} \tilde{n}_s \big) \, , 
\end{align}
\label{eq:ci}
\end{subequations} 
where $\bar{A}$, $\bar{A}_f$ and $\bar{A}_s$ are the mean equatorial moments of inertia of the whole planet, the fluid core and inner core, respectively.  The perturbation in the moment of inertia tensors from external contributions (i.e. due to tidal forces) are written as

\begin{equation}
\tilde{c}^{\,e} = - \bar{A} \frac{\phi_m}{e} {\cal S}_{11} \tilde{\varepsilon}_m \, , \hspace*{0.5cm} \tilde{c}_f^{\,e} = - \bar{A}_f \frac{\phi_m}{e} {\cal S}_{21} \tilde{\varepsilon}_m\, , \hspace*{0.5cm} \tilde{c}_s^{\,e} = - \bar{A}_s \frac{\phi_m}{e} {\cal S}_{31} \tilde{\varepsilon}_m \, , \label{eq:ce}
\end{equation}
where $\phi_m$ is given by Equation (\ref{eq:phims}) below and $e=(C-\bar{A})/\bar{A}$ is the dynamic ellipticity (Equation 3a of D21).

\subsection{The linear system of equations}

Equations (12a-12c) of D21 describe, respectively,  the time rate of change of the angular momenta of the whole of Mercury, the fluid core, and the inner core in the reference frame of the rotating mantle.  Viscoelastic deformations modify these three equations to

\begin{subequations}
\begin{equation}
 ({\omega} - e)\tilde{m} + (1 +{\omega}) \Bigg[ \frac{\bar{A}_f}{\bar{A}} \tilde{m}_f + \frac{\bar{A}_s}{\bar{A}}  \tilde{m}_s + \alpha_3 e_s \frac{\bar{A}_s}{\bar{A}} \tilde{n}_s + \frac{\tilde{c}}{\bar{A}} \Bigg] = \frac{1}{i \Omega_o^2 \bar{A}} \Big(\tilde{\Gamma}_{sun} +\tilde{\Gamma}_{t}  \Big) \, , \label{eq:am1}
 \end{equation}
 
 \begin{equation}
 {\omega} \tilde{m} + \left( 1 + {\omega}  + e_f  \right) \tilde{m}_f -  {\omega} \alpha_1 e_s \frac{\bar{A}_s}{\bar{A}_f} \tilde{n}_s + {\omega} \frac{\tilde{c}_f}{\bar{A}_f} =   \frac{1}{i \Omega_o^2 \bar{A}_f} \Big(- \tilde{\Gamma}_{cmb}  - \tilde{\Gamma}_{icb} \Big) \, , \label{eq:af1}
 \end{equation}
 
 \begin{equation}
( {\omega} - \alpha_3 e_s) \tilde{m} + \alpha_1 e_s \tilde{m}_f +  \left(1+{\omega}\right) \tilde{m}_s  +  \left(1+{\omega} - \alpha_2 \right) \Bigg[ e_s \tilde{n}_s + \frac{\tilde{c}_s}{\bar{A}_s} \Bigg]= \frac{1}{i \Omega_o^2 \bar{A}_s} \Big( \tilde{\Gamma}_{sun}^s  + \tilde{\Gamma}_{ts} + \tilde{\Gamma}_{icb} \Big) \, , \label{eq:as1}  
 \end{equation}
 where $\tilde{\Gamma}_{sun}$, $\tilde{\Gamma}_{sun}^s$ are the gravitational  torques by the Sun on the whole of Mercury and on the inner core alone, respectively, and $\tilde{\Gamma}_{cmb}$, $\tilde{\Gamma}_{icb}$ are the torques from tangential stresses by the fluid core on the mantle at the CMB and on the
inner core at the ICB, respectively.   We have also introduced the torques associated with tidal dissipation (the tidal torque) acting on the whole of Mercury, $\tilde{\Gamma}_{t}$, and on its inner core, $\tilde{\Gamma}_{ts}$; these are developed in section B.4. 
   
The two additional equations of the system are kinematic relations, one that expresses the change in the orientation of the inner core figure as a result of its own rotation, and a second that expresses the invariance of the Laplace pole as seen in the mantle frame.  These are unaffected by deformations and are
 
 \begin{equation}
 \tilde{m}_s  + {\omega}  \tilde{n}_s  = 0 \, , \label{eq:msns}
 \end{equation}
 
\begin{equation}
\tilde{m} + (1+{\omega}) \tilde{\varepsilon}_m  = - (1+{\omega})  \tan I  \, .
 \label{eq:kinI}
 \end{equation}
 \end{subequations}

\subsection{Modification of the solar torque}

For a small mantle obliquity $\tilde{\varepsilon}_m$, the (rigid) gravitational torque by the Sun on the whole of Mercury is given by Equation (14) of D21,

\begin{equation}
\tilde{\Gamma}_{sun}  = -i \Omega_o^2\left[ \bar{A}  \phi_m \,  \tilde{\varepsilon}_m + \bar{A}_s \alpha_3 \phi_s \tilde{n}_s \right]  \, , \label{eq:tqsun}
\end{equation}
where 

\begin{equation}
\phi_m  = \frac{3}{2} \frac{n^2}{\Omega_o^2} \left[ G_{210} \, e + \frac{1}{2} G_{201} \, \gamma  \right] \, , \hspace*{0.5cm}
\phi_s  = \frac{3}{2} \frac{n^2}{\Omega_o^2} \left[ G_{210} \, e_s + \frac{1}{2} G_{201} \, \gamma_s  \right] \, , \label{eq:phims}
\end{equation}
and where $e$, $\gamma$ and $e_s$, $\gamma_s$ are dynamical ellipticities (defined by Equations 3a and 3b of D21), $
G_{210}$ and $G_{201}$ are functions of the orbital eccentricity $e_c$ (defined by Equations 16a and 16b of D21), $n$ is the mean motion and $\Omega_o$ is the rotation frequency.

We adapt Equation (\ref{eq:tqsun}) to include the perturbation in the moment of inertia caused by elastic tidal deformations.  To do so, we follow \cite{baland17}.  Their model does not take into account the misalignment of the inner core tilt (i.e. they assume $\tilde{n}_s=0$).  They write the rigid torque  $\tilde{\Gamma}_{sun}$ as

\begin{equation}
\tilde{\Gamma}_{sun}  = -i \frac{3}{2} n \left( \kappa_{20} + \kappa_{22} \right) \,  \tilde{\varepsilon}_m  \, ,\label{eq:tqsun2}
\end{equation}
where the parameters $\kappa_{20}$ and $\kappa_{22}$ are defined in their Equations (24-25).  The connection between Equations (\ref{eq:tqsun}) and (\ref{eq:tqsun2}) implies that $\frac{3}{2}n (\kappa_{20} + \kappa_{22}) = \Omega_o^2\bar{A} \phi_m$ in our notation.  \cite{baland17} then show how elastic deformations induced by solar tides modify $\kappa_{20}$ and $\kappa_{22}$ (their Equations 53-54), and alter the solar torque to 

\begin{align}
\tilde{\Gamma}_{sun}  & = -i \frac{3}{2} n \left( \kappa_{20} + \kappa_{22} + k_2 MR^2q_t n \left(\frac{1}{6} + \frac{1}{2}e_c^2 + \frac{49}{24} e_c^2  \right)\right) \,  \tilde{\varepsilon}_m  \, ,\nonumber\\
& = -i \frac{3}{2} n \left( \kappa_{20} + \kappa_{22} + k_2 MR^2q_t n \left(\frac{1}{6} +  \frac{61}{24} e_c^2 \right) \right)  \tilde{\varepsilon}_m\, ,
\label{eq:tqsun3}
\end{align}
where $q_t = - 3 R^3 n^2/(G M)$ is a tidal parameter. Substituting $q_t$ and $k_2$ (from Equation \ref{eq:k2Q}) into Equation (\ref{eq:tqsun3}), we get

\begin{equation}
\tilde{\Gamma}_{sun}   = -i \frac{3}{2} n \left( \kappa_{20} + \kappa_{22} - 9 \bar{A} n \frac{n^2}{\Omega_o^2} Re[ {\cal S}_{11} ] \left(\frac{1}{6} +  \frac{61}{24} e_c^2 \right) \right)  \tilde{\varepsilon}_m\, .
\label{eq:tqsun4}
\end{equation}
The difference between Equations (\ref{eq:tqsun4}) and (\ref{eq:tqsun2}) captures the modification of the torque by elastic deformations.  Re-introducing the part of the torque associated with a tilted inner core, and modifying the latter to take into account elastic deformations in the same manner (though it involves the compliance ${\cal S}_{31}$ instead of ${\cal S}_{11}$), we write the modified torque in our notation as  

\begin{equation}
\tilde{\Gamma}_{sun}   = -i \Omega_o^2 \left[ \bar{A} \phi_m^{el}\,  \tilde{\varepsilon}_m + \bar{A}_s \alpha_3 \phi_s^{el} \tilde{n}_s \right]  \, ,\label{eq:tqsun5}
\end{equation}
with 

\begin{subequations}
\begin{equation}
\phi_m^{el}  = \phi_m -  {\cal F}(e_c) Re[{\cal S}_{11}] \, , \hspace*{0.5cm}
\phi_s^{el}  = \phi_s - {\cal F}(e_c) Re[{\cal S}_{31}] \, ,
\label{eq:phimel}
\end{equation}
and where

\begin{equation}
{\cal F}(e_c) = 9 \frac{n^4}{\Omega_o^4} \left(\frac{1}{4} +  \frac{61}{16} e_c^2 \right) \, .
\end{equation}
\end{subequations}

The expression for the torque in Equation (\ref{eq:tqsun5}) includes the effect of elastic deformations associated with the external gravitational potential from the Sun (captured by Equation \ref{eq:ce}).   We further modify the torque to also take into account elastic deformations from internal contributions (captured by Equation \ref{eq:ci}).  For this, we follow section 2.4 of \cite{organowski20} and our final expression of the solar torque is

\begin{equation}
\tilde{\Gamma}_{sun}   = -i \Omega_o^2 \left[ \bar{A} \phi_m^{el}\,  \tilde{\varepsilon}_m + \bar{A}_s \alpha_3 \phi_s^{el} \tilde{n}_s  + \phi_m \frac{\tilde{c}^{i}}{e} + \alpha_s \phi_s \frac{\tilde{c}_s^{i}}{e_s} \right]  \, .\label{eq:tqsun6}
\end{equation}

The solar torque on a rigid inner core is given by Equation (17) of D21.  Following the same procedure as above, elastic deformations modify this torque to 

\begin{equation}
\tilde{\Gamma}_{sun}^s   = -i \Omega_o^2 \left[ \bar{A}_s \alpha_3 \phi_s^{el} (\tilde{\varepsilon}_m + \tilde{n}_s )  + \alpha_s \phi_s \frac{\tilde{c}_s^{i}}{e_s} \right]  \, .\label{eq:tqsuns}
\end{equation}

\subsection{Tidal torque}

We adopt a weak friction tidal model in which the deformed surface of Mercury due to the solar tide matches that based on a purely elastic planet, but delayed by a time lag \cite[][]{darwin1879,alexander73}.
The torque associated with tidal dissipation is \cite[e.g.][Equation 1]{levrard07}, 

\begin{equation}
{\bf \Gamma_t} = 3 \frac{k_2}{Q} \frac{G M_s^2 R^5}{a^6}\left[  \left( f_1 - \frac{f_2 \Omega_o}{2n} {\bf \hat{\Omega}} \cdot  \mitbf{\hat{e}_3^I} \right) \mitbf{\hat{e}_3^I}+ \left( f_1 -\frac{f_2\Omega_o}{2n} \Big(1+ ({\bf \hat{\Omega}} \cdot  \mitbf{\hat{e}_3^I})^2\Big) \right) {\bf \hat{\Omega}} \right]\, , \label{eq:td1}
\end{equation}
where $M_s$ is mass of the Sun, $a$ is the semi-major axis of Mercury's orbit, ${\bf \hat{\Omega}}={\bf \Omega}/\Omega_o$ is the planetary rotation unit vector, and the functions of the eccentricities $f_1$ and $f_2$ are given by

\begin{equation}
 f_1  = \frac{1 + \frac{15}{2} e_c^2 + \frac{45}{8} e_c^4}{(1-e_c^2)^6} \, ,\hspace*{1cm} 
f_2  = \frac{1 + 3 e_c^2 + \frac{3}{8} e_c^4}{(1-e_c^2)^{9/2}} \, .
\label{eq:f1f2}
\end{equation}
Writing $k_2/Q$ in terms of $Im[{\cal S}_{11}]$ using Equation (\ref{eq:k2Q}), and using the definition of the mean motion $n^2 = GM_s/a^3$, $\Omega_o=\frac{3}{2}n$ and ${\bf \hat{\Omega}} \cdot  \mitbf{\hat{e}_3^I}=\cos (Re[\tilde{\varepsilon}_m]) \approx 1$, we can write the tidal torque as

\begin{equation}
{\bf \Gamma_t} = 9 \bar{A} \frac{n^4}{\Omega_o^2} Im[{\cal S}_{11}]\left[  \left( f_1 - \frac{3}{4} f_2 \right) \mitbf{\hat{e}_3^I}  + \left( f_1 - \frac{3}{2} f_2 \right) {\bf \hat{\Omega}} \right] \, . \label{eq:td2}
\end{equation}

We now project this torque onto the equatorial components of the frame attached to Mercury.  If we chose $t=0$ to correspond to when the Cassini plane coincides with the real axis, then with respect to $\mitbf{\hat{e}_3^p}$,  the projection of the $\mitbf{\hat{e}_3^I}$ component of the tidal torque onto the complex plane involves a factor  $-\sin{\tilde\varepsilon_m}\approx -\tilde{\varepsilon}_m$ (see Figure \ref{fig:cassini}b). The part of the torque directed along the rotation vector ${\bf \hat{\Omega}}$ can be divided into a part pointing in the direction of the Laplace pole  $\mitbf{\hat{e}_3^L}$ and a part directed in the Laplace plane.  The former is responsible for a secular change in the orbit of Mercury; as we assume no change in any orbital quantity, we set this part equal to zero.  The remaining part, directed along the Laplace plane, participates in the precession torque.  With the same choice of $t=0$ as above, its projection onto the complex plane of the equator of Mercury involves a factor 

\begin{equation}
\cos (I+\tilde{\varepsilon}_m)\sin(I+\tilde{\varepsilon}_m)\approx \cos I \sin I +(\cos^2 I - \sin^2 I)\tilde{\varepsilon}_m \, .
\end{equation}
Using these projections, the tidal torque is expressed as

\begin{equation}
\tilde{\Gamma}_t = - \Omega_o^2 \bar{A} \, Im[{\cal S}_{11}] \Big[ \phi_m^{t3} \tilde{\varepsilon}_m  +   \phi_m^{t2} \cos I \sin I \Big] \, , \label{eq:td3}
\end{equation}
where 

\begin{equation}
 \phi_m^{t3}  =\Big(\phi_m^{t1} + \phi_m^{t2}\left( \cos^2 I - \sin^2 I\right)\Big)  \, , \hspace*{0.3cm} 
 \phi_m^{t1}  =  9  \frac{n^4}{\Omega_o^4}  \left( f_1 - \frac{3}{4} f_2 \right) \, , \hspace*{0.3cm}  \phi_m^{t2}  =  9  \frac{n^4}{\Omega_o^4}  \left( - f_1 + \frac{3}{2} f_2 \right) \, .
  \label{eq:Phit}
\end{equation}
Truncated to $e_c^2$, we can write

\begin{equation}
\left( f_1- \frac{3}{4} f_2\right)  = \frac{1}{8} \left(2 + {63} e_c^2\right)   \, , \hspace*{1cm} \left(- f_1+ \frac{3}{2} f_2\right)  = \frac{1}{4} \left(2 - 9 e_c^2\right) \, ,
\end{equation}
and the expression for  $\phi_m^{t1}$ and $\phi_m^{t2}$ directly in terms of $e_c$ are

\begin{equation}
 \phi_m^{t1}  =  \frac{9}{4}  \frac{n^4}{\Omega_o^4}  \left(1 + \frac{63}{2} e_c^2 \right) \, , \hspace*{1cm}  \phi_m^{t2}  =  \frac{9}{4}  \frac{n^4}{\Omega_o^4}  \left(2 - 9 e_c^2\right) \, .
  \label{eq:Phit}
\end{equation}

In principle, for a planet with an inner core whose rotation vector is misaligned with that of the mantle, then the deviation from ${\bf \hat{\Omega}}$ within the inner core introduces a correction term to the expression of the torque given by Equation (\ref{eq:td3}).  However, the misalignment of the inner core rotation vector is small and we neglect this correction term.  

The torque on the inner core alone can be constructed in exactly the same manner as that for the whole of Mercury.  The torque has a similar form as that of Equation (\ref{eq:td3}), except it involves the density contrast at the ICB $\alpha_3$ and we must replace $\bar{A}$ with $\bar{A}_s$ and ${\cal S}_{11}$ with ${\cal S}_{31}$:

\begin{equation}
\tilde{\Gamma}_{ts} = - \Omega_o^2 \bar{A}_s \, \alpha_3 Im[{\cal S}_{31}] \Big[ \phi_m^{t3} \tilde{\varepsilon}_m  +   \phi_m^{t2} \cos I \sin I \Big] \, . \label{eq:tdsic}
\end{equation}

\subsection{Modified matrix elements}

The linear system given by Equations (\ref{eq:am1}-\ref{eq:kinI}) can be written in matrix form as $\boldsymbol{\mathsf{M}} \cdot {\bf x} = {\bf y}$ (Equation 22a of D21).  The elements of the vector ${\bf x}$ (Equation 22b of D21) are the 5 unknown rotational variables; solutions for ${\bf x}$ are found by solving this  linear system.  With the addition of elastic deformations, the matrix $\boldsymbol{\mathsf{M}}$ and right-hand side vector ${\bf y}$ given by Equations (22d) and (22c) of D21, respectively, are modified to $\boldsymbol{\mathsf{M}}+\boldsymbol{\mathsf{\delta M}}$ and ${\bf y} + {\bf \delta y}$.  The non-zero elements of $\boldsymbol{\mathsf{\delta M}}$  and ${\bf \delta y}$ are:

\begin{subequations}
\begin{align}
\boldsymbol{\mathsf{\delta M}}_{1,1-3} & = \left(1+{\omega} +\frac{\phi_m}{e} \right) {\cal S}_{1,1-3} + \alpha_3 \frac{\bar{A}_s}{\bar{A}} \frac{\phi_s}{e_s} {\cal S}_{3,1-3} \, ,\\
\boldsymbol{\mathsf{\delta M}}_{1,4} & = \left(1+{\omega} +\frac{\phi_m}{e} \right) {\cal S}_{14} + \alpha_3 \frac{\bar{A}_s}{\bar{A}} \left( \frac{\phi_s}{e_s}  {\cal S}_{34} - {\cal F}(e_c) Re[{\cal S}_{31}] \right) \, ,\\
\boldsymbol{\mathsf{\delta M}}_{1,5} & = -(1+{\omega})\frac{\phi_m}{e} {\cal S}_{11} - {\cal F}(e_c) Re[{\cal S}_{11}] - i \phi_m^{t3} Im[{\cal S}_{11}]\, ,\\
\boldsymbol{\mathsf{\delta M}}_{2,1-4} & = {\omega} {\cal S}_{2,1-4} \, ,\\
\boldsymbol{\mathsf{\delta M}}_{2,5} & = - {\omega} \frac{\phi_m}{e} {\cal S}_{21} \, ,\\
\boldsymbol{\mathsf{\delta M}}_{3,1-3} & = \left(1+{\omega} -\alpha_2 + \alpha_3 \frac{\phi_s}{e_s} \right) {\cal S}_{3,1-3}\, ,\\
\boldsymbol{\mathsf{\delta M}}_{3,4} & = \left(1+{\omega} -\alpha_2 + \alpha_3 \frac{\phi_s}{e_s} \right) {\cal S}_{34} - \alpha_3 {\cal F}(e_c) Re[{\cal S}_{31}] \, ,\\
\boldsymbol{\mathsf{\delta M}}_{3,5} & = -\left(1+{\omega} -\alpha_2 \right)\frac{\phi_m}{e} {\cal S}_{31}  - \alpha_3 {\cal F}(e_c) Re[{\cal S}_{31}] - i \alpha_3 \phi_m^{t3} Im[{\cal S}_{31}]\, , \\
{\bf \delta y}_1 & = i \phi_m^{t2} Im[{\cal S}_{11}] \cos I \sin I \, , \\
{\bf \delta y}_3 & = i \alpha_3 \phi_m^{t2} Im[{\cal S}_{31}] \cos I \sin I \, .
\end{align}
\label{eq:dM}
\end{subequations} 

\section{Computation of the Compliances}

The compliances connected to the misaligned rotation vectors of the whole planet (${\cal S}_{i1}$), of the fluid core (${\cal S}_{i2}$) and of the inner core (${\cal S}_{i3}$) are computed with the standard method presented in many studies \cite[e.g.][]{buffett93,dehant15book}.  

To compute the compliances associated with the inner core tilt (${\cal S}_{i4}$), we follow the method presented in Appendix A of \cite{dumberry08b}.  This method applies for Earth, and it is modified here for Mercury.  A tilt by an angle $\theta_n$ of an elliptical inner core (with geometrical ellipticity ${\epsilon_s}$) produces a radial displacement of degree 2 at the ICB (radius $r_s$) of amplitude $\Delta=r_s \epsilon_s \sin \theta_n$.  Because we use a simplified Mercury model with uniform density in each region, the only perturbation in mass produced by a tilted inner core is at the ICB, a mass load equal to $(\rho_s-\rho_f) \Delta$.  The forcing vector inside the inner core \cite[Equation A16 of][]{dumberry08b} is set to zero.  We then model the viscoelastic response of a reference spherical planet to this degree 2 mass load at the ICB.  Written in terms of the standard set of 6 linear variables $y_{1-6}$ \cite[see their definitions in][]{dumberry08b}, the mass load boundary conditions at the ICB are 

\begin{subequations}
\begin{align}
y_1^s & = -\frac{y_5^f}{g} + A_1 \, ,  \\
y_2^s & = A_1 \rho_f g - (\rho_s -\rho_f)g \Delta \, ,  \\
y_3^s & = A_2 \, ,  \\
y_4^s & =0 \, , \\
y_5^s & = y_5^f \, ,  \\
y_6^s & = y_6^f + 4\pi G \rho_f A_1 - 4\pi G (\rho_s -\rho_f) \Delta \, , 
\end{align}
\end{subequations}
where $g$ is the gravitational acceleration, superscripts $s$ and $f$ denote quantities on the solid and fluid side of the ICB, respectively, and $A_1$ and $A_2$ are constants of integration.  Five more constants of integrations are introduced by the boundary conditions at the centre and at the CMB \cite[unchanged from those used in][]{dumberry08b}.  Solutions for the viscoelastic deformations of the whole planet are found for an assumed (non-dimensional) radial displacement equal to 1, and the compliances ${\cal S}_{14}$, ${\cal S}_{24}$ and ${\cal S}_{34}$ are then computed from the perturbation in the moments of inertia of the whole planet, the fluid core and the inner core, respectively.

The numerical values of all compliances depend on the choice of a reference model of density and viscoelastic parameters (the Lam\'e parameter $\lambda$ and shear modulus $\mu$) as a function of radius.  We assume uniform values in each of the inner core, the fluid core, the mantle and the crust.  The density of the crust is taken as 2974 kg m$^{-3}$, that of the inner core as 8800 kg m$^{-3}$.  The densities of the fluid core and mantle depend on inner core size and are specified by the method detailed in section 3.1 of D21.   In the crust, mantle and inner core, the moduli $\lambda$ and $\mu$ are frequency dependent.  We assume a Maxwell rheology, in which $\lambda$ and $\mu$ depend on the viscosity $\eta$ and the frequency of the deformation $\omega'$ through \cite[e.g.][]{wu82}

\begin{equation}
    \lambda = \frac{( i \omega' \lambda_o + \frac{\kappa}{\eta} \mu_o )}{(i \omega' + \frac{1}{\eta} \mu_o)} \, , \quad  \mu = \frac{ i \omega' \mu_o}{( i \omega' + \frac{1}{\eta} \mu_o )} \, ,\label{eq:lambdamu} 
\end{equation}
where $\lambda_o$ and $\mu_o$ denote the moduli in the elastic limit ($\omega' \gg \mu_o/\eta$) and $\kappa= \lambda_o + \frac{2}{3} \mu_o$ is the bulk modulus.  For deformations connected to the Cassini state, the forcing frequency is $\omega' = \omega \Omega_o$, where $\omega$ is given by Equation (\ref{eq:omega1}) and $\Omega_o=2\pi/58.64623$ day$^{-1}$ is the sidereal frequency.  $\lambda_o$ and $\mu_o$ are specified in terms of uniform compressional ($V_p$) and shear ($V_s$) seismic wave velocities and density $\rho$ within each region. They are computed from, 

\begin{equation}
\mu_o = \rho V_s^2 \, , \hspace*{1cm} \lambda_o = \rho V_p^2 - 2 \mu_o \, .
\end{equation}
In doing so, we make the implicit assumption that the timescale of propagation of seismic waves within the solid regions of Mercury is sufficiently short that deformations are in the elastic limit.  The $V_p$ and $V_s$ values that we use are listed in Table \ref{tab:seis} and are based on those presented in \citet{rivoldini09,rivoldini11}, except for $V_s$ in the mantle and crust.  The common numerical value of the latter two is computed by ensuring that, for each choice of inner core size, for chosen values of the viscosity in each of the solid regions, and with $\omega'= \omega \Omega_o$, the $\mu$ and $\lambda$ values that are calculated via Equation (\ref{eq:lambdamu}) yield a second degree tidal Love number $k_2$ which is equal to 0.55.  This ensures that all interior models that we consider in our study are consistent with recent observations of tidal deformations \cite[][]{konopliv20,genova19}.  Note that the observed value of $k_2=0.55$ is based on sectorial tides whose frequency is equal to the mean motion $n=2\pi/87.96935$ day$^{-1}$.  Our computation is carried instead at a frequency close to $\Omega_o$, so in effect we make the assumption that $k_2\approx 0.55$ also at a frequency of $\Omega_o$.

\begin{table}[]
\begin{tabular}{lllll}
\hline
Seismic parameter & Crust & Mantle & Fluid core & Inner core \\ \hline
$V_p$ (m s$^{-1}$) & 8000 & 8500 & 5000 & 7000 \\
$V_s$ (m s$^{-1}$) & calculated & calculated & 0 & 3800 \\
$\rho$ (kg m$^{-3}$) & 2974 & calculated  & calculated & 8800 \\
\hline
\end{tabular}
\caption{\label{tab:seis} Seismological parameters used in our calculations.  $V_p$ and $V_s$ are, respectively, the compressional and shear seismic velocities.  The density ($\rho$) for the mantle and fluid core and the shear seismic wave ($V_s$) for the mantle and crust depend on inner core size. }
\end{table}

\begin{figure}
\begin{center}
    \includegraphics[height=6cm]{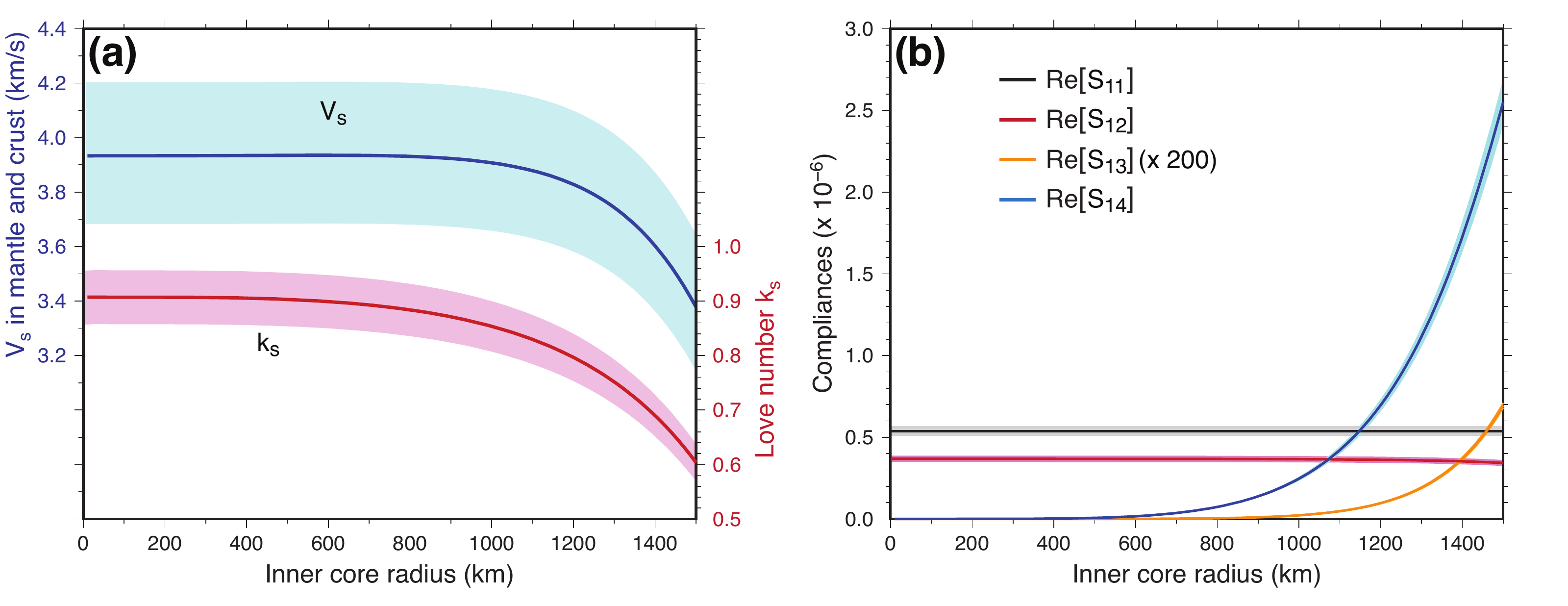}  
    \caption{\label{fig:comp} (a) Shear wave seismic velocity $V_s$ in the mantle and crust  (blue), the real part of the Love number $k_s$ (red) and (b) the real parts of the compliances ${\cal S}_{11}$ (black), ${\cal S}_{12}$ (red), ${\cal S}_{13}$ (orange), ${\cal S}_{14}$ (blue) as a function of inner core radius.  The viscosity is set to $10^{20}$ Pa s in all solid regions.  In both panels, the solid lines for each variables are for a Mercury model with $k_2$ set equal to 0.55, and the coloured shaded areas bracket a range of $k_2$ between 0.52 and 0.58. }
\end{center}
\end{figure}

Figure \ref{fig:comp}a shows an example of how the seismic shear wave velocity $V_s$ in the mantle and crust changes as a function of inner core size ($r_s$) in order to match $k_2=0.55$.  This is for a calculation where the viscosity in the crust, mantle and inner core is set to $\eta = 10^{20}$ Pa s; with this choice, deformations in the solid regions are firmly in the elastic limit.  $V_s$ is modified from 3.93 km s$^{-1}$ for a small or no inner core, to 3.37 km s$^{-1}$ for $r_s=1500$ km. We also show on Figure \ref{fig:comp} how $V_s$ is modified for a range of $k_2$ values between $0.52-0.58$ (for the same viscosity $\eta = 10^{20}$ Pa s in all solid regions).  For other choices of viscosity, for instance a lower value in the mantle, the profile of $V_s$ as a function of inner core size would be modified, as then a different value of $\mu_o$ is required in order to match $k_2=0.55$.

Since $\lambda$ and $\mu$ calculated from Equation (\ref{eq:lambdamu}) are complex, the compliances ${\cal S}_{ij}$ are also complex.  Their real parts capture the deformations that are in-phase with the applied forcing and their imaginary parts, those that are out-of-phase by a quarter of a cycle.  Figure \ref{fig:comp}b shows how the real parts of the compliances ${\cal S}_{1,1-4}$ change as a function of $r_s$. These are the four compliances that have the largest influence on the Cassini state solution.  As we enforce $k_2$ to remain fixed at 0.55, regardless of inner core size, $Re[{\cal S}_{11}]$ (connected to $k_2$ through Equation \ref{eq:k2Q}) also remains constant and is equal to $5.38 \times 10^{-7}$.  $Re[{\cal S}_{12}]$ is reduced slightly from $3.68 \times 10^{-7}$ for a small inner core to $3.43 \times 10^{-7}$ for a large inner core.  The two compliances connected to the inner core, ${\cal S}_{13}$ and ${\cal S}_{14}$, are both very small for a small inner core and increase substantially with inner core size.  $Re[{\cal S}_{13}]$ remains small in amplitude; it is multiplied by a factor 200 on Figure \ref{fig:comp}b so as to be visible and its maximum value is $3.47 \times 10^{-9}$ for $r_s=1500$ km.   $Re[{\cal S}_{14}]$ becomes larger than both $Re[{\cal S}_{11}]$ and $Re[{\cal S}_{12}]$ once $r_s > 1150$ km, reaching an amplitude of $2.55 \times 10^{-6}$ for $r_s=1500$ km. We also show on Figure \ref{fig:comp} how $Re[{\cal S}_{1,1-4}]$ are modified for a range of $k_2$ values between $0.52-0.58$. The important point to note is that choosing a different reference $k_2$ value does not induce a large change in these compliances; the choice of inner core size has a much larger effect on $Re[{\cal S}_{13}]$ and $Re[{\cal S}_{14}]$.

\begin{figure}
\begin{center}
    \includegraphics[height=6cm]{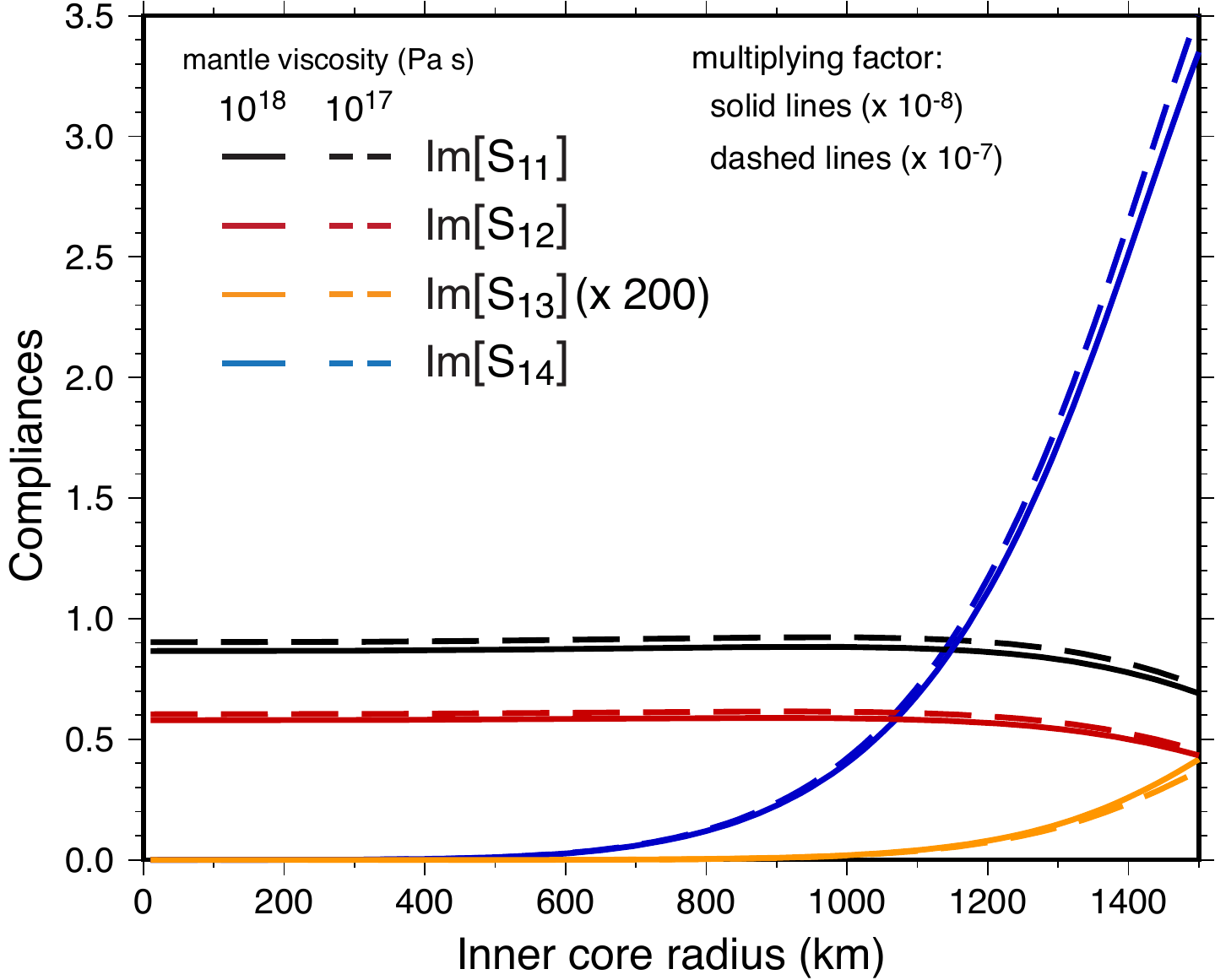}  
    \caption{\label{fig:compimag} The imaginary parts of the compliances ${\cal S}_{11}$ (black), ${\cal S}_{12}$ (red), ${\cal S}_{13}$ (orange), ${\cal S}_{14}$ (blue) as a function of inner core radius for two choices of mantle viscosity: $10^{18}$ Pa s (solid lines) and $10^{17}$ Pa s (dashed lines).  Numerical values must be multiplied by a factor $10^{-8}$ for solid lines, and $10^{-7}$ for dashed lines. The viscosity of the inner core and crust is set to $10^{20}$ Pa s.}
\end{center}
\end{figure}

Because each of our interior model is constrained to match $k_2=0.55$, the real parts of ${\cal S}_{1,1-4}$ do not change when the viscosity of the mantle and/or inner core is reduced.  The imaginary parts of ${\cal S}_{1,1-4}$, however, increase in amplitude when the viscosity of the mantle is reduced. Figure \ref{fig:compimag} shows how they change as a function of $r_s$ for two different choices of mantle viscosity, $10^{18}$ and $10^{17}$ Pa s.  The imaginary parts of ${\cal S}_{1,1-4}$ vary with $r_s$ in a way which is similar to their real parts.  Their amplitudes increase in proportion with the decrease in mantle viscosity.  The quality factor $Q$ is connected to ${\cal S}_{11}$ by $Q=Re[{\cal S}_{11}]/ Im[{\cal S}_{11}]$ (see Equation \ref{eq:k2Q}); a reduction in mantle viscosity  leads to an increase in $Im[{\cal S}_{11}]$ and to a lower $Q$. 

The perturbation in the moment of inertia tensor of the whole planet caused by a tilt of an elliptical rigid inner core with dynamical ellipticity $e_s$ is $\bar{A}_s \alpha_3 e_s \tilde{n}_s$. The additional perturbation caused by global deformations is $\bar{A} ({\cal S}_{13} + {\cal S}_{14}) \tilde{n}_s$ (from Equation \ref{eq:ci1} and using $\tilde{m}_s = \tilde{n}_s$). Since ${\cal S}_{13} \ll {\cal S}_{14}$, we can approximate the total moment of inertia perturbation induced by an inner core tilt as, 

\begin{equation}
\left( \bar{A}_s \alpha_3 e_s + \bar{A} {\cal S}_{14} \right)  \tilde{n}_s  =  \bar{A}_s \alpha_3 e_s  \left(1 + k_s\right) \tilde{n}_s \, ,
\end{equation}
where the Love number $k_s$ is given by

\begin{equation}
k_s = \frac{\bar{A}}{\bar{A}_s} \frac{{\cal S}_{14}}{\alpha_3 e_s} \, .
\end{equation}
We show in Figure \ref{fig:comp}a how the real part of $k_s$ varies with $r_s$. The Love number $k_s$ is of order 1 and provides a convenient way to add the contribution from deformations to the change in the moment of inertia of the whole planet caused by a tilted inner core, as it was done in the case of Earth in \cite{dumberry08b}.

A Maxwell model is likely not a very accurate representation of the rheology of Mercury's mantle. A better choice would be to use an Andrade-pseudoperiod model \cite[e.g.][]{padovan14,steinbrugge18}.  Our choice is instead one of convenience. A Maxwell model provides a simple way to characterize viscoelastic deformations directly in terms of viscosity values, thus limiting the number of model parameters.  Furthermore, a Maxwell model is also straightforward to incorporate in the framework of our rotational model; the same strategy was used in previous studies using the same framework  \cite[e.g.][]{greff00,koot11,organowski20}.  Our primary goal is to recover a first order connection between the phase lag angle and the bulk viscosities of the mantle and inner core.  As we are focused on one single frequency, that associated with the Cassini state, assuming a Maxwell model is sufficient to accomplish this task. Moreover, because we assume uniform material properties in the mantle, instead of taking into account their radial variations, the viscosity that we recover represents a bulk value averaged over the entire mantle, so it can be regarded at best as an order of magnitude estimate.  In this spirit, using a Maxwell model rather than a more accurate rheology is sufficient, although we need to remain alert to the fact that the viscosity values that we recover do depend on this choice.  As an example, a given rigidity is achieved with a higher viscosity in an Andrade rheology compared to a Maxwell rheology \cite[e.g][]{padovan14}. In order to obtain the same tidal quality factor $Q$ -- the parameter ultimately tied to the mantle phase lag -- the mantle viscosity would need to be larger in an Andrade model.

 \acknowledgments
Comments and suggestions by Nicolas Rambaux and an anonymous referee helped to improve the clarity of this work.  Some figures were created using the GMT software \cite[]{gmt}. The source codes, data files and scripts to reproduce all figures are freely accessible at \cite{macpherson22data}. This work was supported by a NSERC/CRSNG Discovery Grant.    


\end{document}